\documentclass{article}
\usepackage{times}
\usepackage{epsfig}
 
\makeatletter

\DeclareSymbolFont{boldmath}{OML}{cmm}{b}{it}
\DeclareSymbolFontAlphabet{\mathb}{boldmath}
\DeclareMathAlphabet{\mathbb}{U}{msb}{m}{n}
\DeclareMathAlphabet{\mathfrak}{U}{euf}{b}{n}

\DeclareMathSymbol{\balpha}{0}{boldmath}{"0B}
\DeclareMathSymbol{\bbeta}{0}{boldmath}{"0C}
\DeclareMathSymbol{\bgamma}{0}{boldmath}{"0D}
\DeclareMathSymbol{\bdelta}{0}{boldmath}{"0E}
\DeclareMathSymbol{\bepsilon}{0}{boldmath}{"0F}
\DeclareMathSymbol{\bzeta}{0}{boldmath}{"10}
\DeclareMathSymbol{\bfeta}{0}{boldmath}{"11}
\DeclareMathSymbol{\btheta}{0}{boldmath}{"12}
\DeclareMathSymbol{\biota}{0}{boldmath}{"13}
\DeclareMathSymbol{\bkappa}{0}{boldmath}{"14}
\DeclareMathSymbol{\blambda}{0}{boldmath}{"15}
\DeclareMathSymbol{\bmu}{0}{boldmath}{"16}
\DeclareMathSymbol{\bnu}{0}{boldmath}{"17}
\DeclareMathSymbol{\bxi}{0}{boldmath}{"18}
\DeclareMathSymbol{\bpi}{0}{boldmath}{"19}
\DeclareMathSymbol{\brho}{0}{boldmath}{"1A}
\DeclareMathSymbol{\bsigma}{0}{boldmath}{"1B}
\DeclareMathSymbol{\btau}{0}{boldmath}{"1C}
\DeclareMathSymbol{\bupsilon}{0}{boldmath}{"1D}
\DeclareMathSymbol{\bphi}{0}{boldmath}{"1E}
\DeclareMathSymbol{\bchi}{0}{boldmath}{"1F}
\DeclareMathSymbol{\bpsi}{0}{boldmath}{"20}
\DeclareMathSymbol{\bomega}{0}{boldmath}{"21}
\DeclareMathSymbol{\beps}{0}{boldmath}{"22}
\DeclareMathSymbol{\bthet}{0}{boldmath}{"23}
\DeclareMathSymbol{\bomeg}{0}{boldmath}{"24}
\DeclareMathSymbol{\bvphi}{0}{boldmath}{"27}

\def\beq{\arraycolsep=2pt\begin{eqnarray}\@ifnextchar[{\eql}{}}
\def\eeq{\end{eqnarray}}
\def\eql[#1]{\label{#1}}
\def\nwl{\nonumber\\}
\def\aln#1{& #1 &}
\def\fl{\hspace{-1em}}
\def\txt#1{\quad\hbox{#1}\quad}
\def\theequation{\thesection.\arabic{equation}}
\@addtoreset{equation}{section}

\def\subsec#1{\subsubsection*{#1}}

\def\sref#1{section~\ref{#1}}
\def\fref#1{figure~\ref{#1}}

\def\eref#1{(\ref{#1})}

\def\ssty{\scriptstyle}

\def\^{{}^}
\def\_{{}_}

\def\follows{\quad\Rightarrow\quad}
\def\equivalent{\qquad\Leftrightarrow\qquad}
\def\case#1#2{{\textstyle{{#1}\over{#2}}}}
\def\intl#1#2{\int_{\makebox[0pt][l]{$\ssty #1$}}
                  ^{\makebox[0pt][l]{$\ssty #2$}}\,}
\def\inti#1{\int_{\makebox[0pt][l]{$\ssty #1$}}\,}
\def\opr#1{\widehat #1}
\def\Pexp{\mathcal{P}\!\exp}
\def\expo#1{e^{#1}}
\def\expt#1{e^{#1}}
\def\comm#1#2{\big[\opr #1,\opr #2\big]}
\def\pois#1#2{\big\{#1,#2\big\}}

\def\d{\mathrm{d}}
\def\del{\partial}
\def\deldel#1{\frac{\del}{\del #1}}
\def\dd{\mathrm{d}}
\def\i{\mathrm{i}}
\def\Trr#1{\mathrm{Tr}(#1)}
\def\Trrr#1{\mathrm{Tr}\big(#1\big)}
\def\inv{\!\!^{-1}}
\def\ket#1{| #1 \rangle}

\def\braket#1#2{\langle #1 | #2 \rangle}
\def\Ket#1{\,\big| #1 \big\rangle}
\def\Bra#1{\big\langle #1 \big|\,}
\def\Braket#1#2{\big\langle #1 \,\big|\, #2 \big\rangle}
\def\norm#1{\big\|\, #1 \,\big\|}
\def\expect#1{\langle #1 \rangle}
\def\pBraket#1#2{\big\langle #1 \,\big|\, #2 \big\rangle_{\mathrm{phys}}}

\def\grpO{\mathsf{O}}
\def\grpSL{\mathsf{SL}}
\def\grpSO{\mathsf{SO}}
\def\grpSU{\mathsf{SU}}
\def\grpISL{\mathsf{ISL}}
\def\grpISO{\mathsf{ISO}}
\def\algsl{\mathfrak{sl}}
\def\algso{\mathfrak{so}}

\def\Zset{\mathbb{Z}}
\def\Rset{\mathbb{R}}
\def\Mset{\mathbb{M}}   

\def\tang{\mathsf{T}}

\def\spce{\mathcal{N}}
\def\sptm{\mathcal{M}}   
\def\sphr{\mathcal{S}} 
\def\sbst{\mathcal{U}}

\def\phsp{\mathcal{P}}
\def\exps{\widetilde\phsp}
\def\hlsp{\mathcal{H}}
\def\phhs{\hlsp_\mathrm{phys}}

\def\Lvec{\mathcal{L}}
\def\Rvec{\mathcal{R}}

\def\gam{\bgamma}
\def\one{\mathbf{1}}
\def\eee{\mathb{e}} 
\def\omg{\bomega}
\def\covD{D}
\def\flds{\mathb{F}}
\def\dete{e}
\def\fff{\mathb{f}} \def\fffv{f}  
\def\ggg{\mathb{g}} 
\def\llpar{\mathb{k}} 
      
\def\llgen{\blambda}  \def\llgenv{\lambda}  
\def\ltgen{\bxi}           
\def\dtgen{\mathb{n}} \def\dtgenv{n} 
\def\rlpar{\mathb{h}}   
\def\rtpar{\mathb{n}}  
\def\rlgen{\mathb{c}}  
\def\rtgen{\mathb{d}}  
\def\dfgen{\xi}        
\def\msgen{\eta}      
\def\mom{\mathb{p}} \def\momv{p}  
\def\moms{u} \def\smom{{\vec p}}
\def\hol{\mathb{u}} \def\holv{u} 
\def\ang{\mathb{v}} \def\angv{J} 
\def\xpos{\mathb{x}} \def\xposv{x}  
\def\ypos{\mathb{y}} \def\yposv{y} 
 \def\qposv{q} 
 \def\dposv{d} 
\def\avec{\mathb{v}} \def\avecv{v}
\def\bvec{\mathb{w}} \def\bvecv{w}
\def\chol{\mathb{U}}
\def\min{\mathrm{min}}

\def\diag{\mathrm{diag}}
\def\mod{\mathrm{mod\,}}
\def\eps{\varepsilon}
\def\p{\varphi}
\def\mult{\zeta}   
\def\xmult{\bchi}                      
\def\mass{m}  
\def\defangle{\alpha}
\def\anti{\epsilon}                     
\def\rlim{R}

\def\lag{L}
\def\lagint{\lag_\mathrm{int}}
\def\lagbnd{\lag_\mathrm{bnd}}
\def\lagprt{\lag_\mathrm{prt}} 
\def\lagkin{\lag_\mathrm{kin}}
\def\ham{H}
\def\con{C}
\def\pot{\Theta}
\def\sym{\Omega}

\def\contract{\cdot}
\def\config{\Phi}
\def\ads{\mathcal{G}} 
\def\newton{G}
\def\lpl{\ell_{\mathrm{P}}}
\def\unit{U}  
\def\euch{\chi}
\def\euph{\phi}
\def\eurh{\rho}

\def\TTT{T}   
\def\ttt{t}   
\def\SSS{S^2} 
\def\sss{s^2} 
\def\RRR{R^2}  
\def\sdis{l}       
\def\scon{\lambda} 
\def\MMM{M}    
\def\mmm{n}   

\def\laplace{\Delta}

\makeatother

\begin{document}

\begin{flushright}
THU-97/22\\gr-qc/9708054
\end{flushright}

\LARGE
\begin{center}
  {\bf Quantum Mechanics of a Point Particle in \\ 2+1 Dimensional Gravity}
\end{center}
\normalsize

\vspace{1ex}

\begin{center}
     Hans-J\"urgen Matschull\footnote{E-mail: H.J.Matschull@fys.ruu.nl} and
     Max Welling\footnote{E-mail: welling@vision.caltech.edu}\\
     Instituut voor Theoretische Fysica, Universiteit Utrecht\\
     Princetonplein 5, 3584 CC Utrecht, The Netherlands
\end{center}

\vspace{1ex}

\begin{abstract}
We study the phase space structure and the quantization of a pointlike particle in 2+1 dimensional gravity. By adding boundary terms to the first order Einstein Hilbert action, and removing all redundant gauge degrees of freedom, we arrive at a reduced action for a gravitating particle in 2+1 dimensions, which is invariant under Lorentz transformations and a group of generalized translations. The momentum space of the particle turns out to be the group manifold $\grpSL(2)$. Its position coordinates have non-vanishing Poisson brackets, resulting in a non-commutative quantum spacetime. We use the representation theory of $\grpSL(2)$ to investigate its structure. We find a discretization of time, and some semi-discrete structure of space. An uncertainty relation forbids a fully localized particle. The quantum dynamics is described by a discretized Klein Gordon equation.
\end{abstract}

\vspace{1ex}

\section*{Outline}

The most important property of general relativity in 2+1 dimensions is the lack of local degrees of freedom \cite{Star,DJH}. There are not enough dimensions to support gravitational waves, which transmit the gravitational interaction. The theory is not completely ``void'' however. One can introduce so called topological degrees of freedom by choosing a spacetime manifold that is not simply connected. The study of these cosmological models was greatly facilitated by the observation that 2+1-dimensional gravity is equivalent to a Chern-Simons theory of the Poincar\'e group \cite{Achucarro,Witten}. The physical degrees of freedom can be described by Wilson loops. Also a great deal of progress was made in the quantization of these models, although it is still not clear whether all approaches result in equivalent quantum theories \cite{Carlip}.

Another group of degrees of freedom are usually introduced as punctures in the surface. To these punctures one can attach a parameter, the mass, by adding a particle Lagrangian to the Einstein Hilbert action. It consists of a kinetic term and a constraint, which puts the particle on a mass shell. Because of the topological nature of the interaction, the curvature vanishes everywhere except at the location of the particle. The effect of the mass is to cut out a wedge from spacetime in such a way that the particle is located at the tip of a cone. For a static particle for instance, the deficit angle of this cone is $\defangle = 8\pi\newton\,\mass$, where $\mass$ is the mass of the particle. Newtons constant $\newton$ has the dimension of an inverse mass in three dimensions. Multi particle models are well understood at the level of classical equations of motion \cite{Hooft1}. At the quantum level, it is possible to calculate the exact scattering amplitude for two gravitating particles \cite{Hooft0}, but a
consistent multi particle Hilbert space has not been found.

The fact that point particles in 2+1 dimensional gravity are non-trivial objects was demonstrated only recently by 't~Hooft \cite{Hooft4}. He suggested that the momentum of the particle, which is canonically conjugate to the distance from an arbitrary origin, is actually an angle. This fact was exploited to perform an unusual quantization. He defined the momentum on a sphere and the energy on a circle, resulting in non-commuting operators for the coordinates. The Schr\"odinger equation became a Dirac equation on some kind of spacetime lattice. This was not the first time that someone proposed to use a curved momentum space however. Back in 1946 Snyder \cite{Snyder} proposed to use de-Sitter or anti-de-Sitter space as momentum space. In any case, the deviation from flat momentum space will not be noticeable at small energy scales. However, at large energy scales, the deviation influences the short distance behaviour in spacetime.  

In this article, we want to derive the phase space of a gravitating particle from first principles. We start from the first order Einstein Hilbert action for 2+1 dimensional gravity and add a particle term to it, which is chosen such that the neighbourhood of the particle looks like a cone with a deficit angle of $8\pi\newton\,\mass$. We also have to add a boundary term to the action, in order to render the Lagrangian finite and to obtain the correct equations of motion at spatial infinity. This procedure will, almost automatically, introduce an additional degree of freedom into the system, which can be associated with an observer sitting at infinity. As a result, the gravitational field, which itself has no local degrees of freedom, describes the interaction of the particle with an external observer.

After dividing out the gauge degrees of freedom, we arrive at a finite dimensional reduced phase space, which is spanned by position and momentum coordinates. The position is a vector in Minkowski space, which can be interpreted as the location of the particle measured in the rest frame of the external observer. So far, the description is the same as that of a relativistic point particle in flat Minkowski space. However, the momentum of the particle is not a vector, but an element of the spinor representation of the Lorentz group $\grpSL(2)$. So, the momentum space is indeed curved, but it is not a sphere. Instead, it is locally isomorphic to anti-de-Sitter space. Its cotangent space is the algebra $\algsl(2)$, which is isomorphic to Minkowski space, and thus to the position space of the particle. Hence, all together the phase space becomes the cotangent bundle $\tang_*\grpSL(2)$.

This has some consequences for the Poisson algebra and the symmetries of the phase space. For example, the components of the position vector will not commute with each other. At the quantum level, this leads to a new kind of uncertainty relation, which states that it is not possible to localize the particle within a region that is smaller than the Planck size. We can also say that the quantized particle lives in a ``non-commutative'' spacetime. It is important to note that this comes out as a result. We do not start from a non-commutative structure of spacetime or anything like that \emph{a priori}. To see what such a non-commutative spacetime looks like, we set up a quantum representation that is very similar to the Euclidean angular momentum Hilbert space in three dimensions.

The basic symmetry group that acts on the reduced phase space, and then also on the quantum Hilbert space, turns out to be the Poincar\'e group. This is not unexpected, because the position coordinate of the particle lives in Minkowski space. However, and this is a somewhat surprising effect, this group is not represented in the ordinary way. In particular, a translation is not represented as a simple shift of the position coordinates, but as a more complicated deformed translation, which also involves the momenta. Only for small momenta, it reduces to an ordinary translation. That the Poincar\'e group naturally acts on the reduced phase space can also be seen from the fact that the latter is, as a manifold, a special representation thereof, $\tang_*\grpSL(2) \simeq \grpSL(2) \times \algsl(2) \simeq \grpISL(2)$, and the natural action is the adjoint action of this group onto itself.

The absence of the ordinary translation symmetry has some consequences for the quantum theory as well. It turns out that there is a distinguished point in spacetime. The light cone emerging from this point divides the quantum spacetime into regions with quite different structures. Space and time are discrete in the causal past and future of this point, whereas in the region that is spacelikely separated from this point, space is continuous but time is still discrete. Moreover, there is a smallest distance to this origin, in the sense that the particle cannot be localized in a region of spacetime that is closer than one Planck length from the origin.

Another peculiar effect is that spacetime has an inherent orientation. At the classical level, this shows up in the Poisson brackets of the position coordinates, which involves a Levi Civita symbol. At the quantum level, it manifests itself in the fact that, if we try to localize the particle in a small region, then we have to give it a certain angular momentum as well, which has to point into a fixed direction. Both the distinguished origin of spacetime and its orientation can be considered as parameters of the theory that have to be determined ``by experiments''. They can take any value, but it is not possible to remove them completely.

The dynamics of the particle can be described by a mass shell constraint, which is very similar to that of a relativistic point particle. At the quantum level, it becomes a kind of Klein Gordon equation, which is however not a second order differential equation, but a second order difference equation. This is due to the discretization of time, which implies that the time evolution equation has to be a step equation. In the limit where the Planck length goes to zero, it becomes the usual Klein Gordon equation. This is actually true for everything derived from the reduced phase space structure. If we insert the correct physical units, and then switch off the gravitation by making the Planck length (or Newtons constant) small, we recover the usual description of a relativistic point particle.

All these unfamiliar effects can therefore be thought of as gravitational corrections to the relativistic point particle and its quantization, and they result from the combination of quantum mechanics and general relativity. We are aware that this is just a one particle model and the hope is that in a more sophisticated, multi particle model, some of the details will become somewhat clearer. For example, the origin of spacetime might be removed and the absolute position vector of the particle, refering to this origin, be replaced by the relative position between two particles. The shortest distance would then act as a natural cutoff of the theory. Some progress has been made towards this goal at the level of the classical phase space \cite{Matschull}, and a simple model has been worked out in the quantum case \cite{Welling}, but a full understanding is still lacking.

\section{The spinor representation of 2+1 gravity} 
\label{dreibein}

For the purpose of this article, it is most convenient to use the dreibein representation of Einstein gravity. The basic field variables are then the dreibein $e_\mu\^a$ and the $\algso(1,2)$ spin connection $\omega_{\mu ab}=-\omega_{\mu ba}$. Here, greek indices $\mu,\nu,\dots$ are tangent indices of the spacetime manifold $\sptm$, and latin indices $a,b,\dots$, taking the values $0,1,2$, are three dimensional Minkowski space indices, transforming under the vector representation of the Lorentz group $\grpSO(1,2)$. They are raised and lowered by the metric $\eta_{ab}=\diag(-1,1,1)$, and the Levi-Civita symbol $\eps^{abc}$ is defined such that $\eps^{012}=1$.

A brief introduction into notations and conventions, the structure of the Lorentz group, its spinor representation $\grpSL(2)$, and the associated Lie algebras is given in \ref{lorentz}. A special feature of the three dimensional Lorentz algebra is that it is, as a vector space, isomorphic to Minkowski space itself, and thus the vector representation of the Lorentz algebra is equivalent to its adjoint representation. We can use this to convert the dreibein as well as the spin connection into $\algsl(2)$ valued one-forms, defined by
\beq
  \eee_\mu = e_\mu\^a \, \gam_a, \qquad
  \omg_\mu = \case14 \omega_{\mu ab} \, \gam^a \gam^b.
\eeq
The gamma matrices are those introduced in \eref{gamma}, and bold letters are used to denote $2\times2$ matrices. As a general convention, we shall always use the same symbols, either with vector indices attached or in bold face, to denote the same physical quantity in different representations.

\subsec{Local frames}

The dreibein can be considered as a map from the tangent bundle of the spacetime manifold $\sptm$ into a Minkowski space, or $\algsl(2)$ bundle over $\sptm$. A tangent vector $\avecv^\mu(x)$ at some point $x\in\sptm$ is mapped onto a vector $\avec(x) = \avecv^\mu(x)\eee_\mu(x)$ in a ``local Lorentz frame'', or simply a ``local frame'' at $x$. The scalar product of two such vectors is given by
\beq[metric]
  \case12 \Trr{\avec \bvec} 
  = \case12 \Trr{\eee_\mu\eee_\nu}  \, \avecv^\mu \, \bvecv^\nu
  = g_{\mu\nu} \, \avecv^\mu \, \bvecv^\nu.
\eeq
This defines the spacetime metric $g_{\mu\nu}$, which is required to be invertible. Equivalently, the dreibein determinant has to be non-zero, and without loss of generality we can restrict it to be positive,
\beq[det]
   \dete = \case1{12} \eps^{\mu\nu\rho} \, 
                  \Trr{\eee_\mu\eee_\nu\eee_\rho} > 0.
\eeq
This condition implies that the dreibein provides an isomorphism between the tangent bundle of $\sptm$ and the local frame bundle. So, in order to get an invertible metric, the latter has to be isomorphic to the tangent bundle of the given manifold $\sptm$. We shall see that, for our special model, the tangent bundle of the spacetime manifold is trivial. Let us therefore in the following assume that this is the case. The bundle of local frames is then also trivial, and we can consider the dreibein as an $\algsl(2)$ valued one-form on $\sptm$. The same holds for the spin connection, which defines the covariant derivative of vectors in the local frames,
\beq[cov-div]
  \covD_\mu \avec = \del_\mu \avec + [\omg_\mu,\avec].
\eeq
From this, we derive the following useful quantities. The field strength is defined by
\beq[FF-def]
  [\covD_\mu,\covD_\nu] \avec = [\flds_{\mu\nu},\avec]
  \follows 
  \flds_{\mu\nu} = \del_\mu \omg_\nu - \del_\nu \omg_\mu +
                            [\omg_\mu,\omg_\nu].
\eeq
Moreover, given a curve in $\sptm$, connecting two points $x$ and $y$, we define the ``transport operator''
\beq[hol-def]
  \chol(x,y) = \Pexp \intl xy \d s \, \omg_s,
\eeq
where $\omg_s$ is the pullback of $\omg_\mu$ on the curve. The path ordering is such that factors corresponding to lower parameters $s$, that is, closer to $x$, appear to the left. The matrix $\chol(x,y)\in\grpSL(2)$ describes the parallel transport from the local frame at $x$ along the given curve into the local frame at $y$. A vector $\avec(x)$ is mapped onto $\avec(y)=\chol(y,x)\avec(x)\chol(x,y)$, where $\chol(y,x)=\chol(x,y)^{-1}$. Note that, at this point, it is important that the tangent bundle of $\sptm$ is trivial. Otherwise we had to introduce transition functions into the definition \eref{hol-def}, whenever the curve enters a new coordinate chart.

\subsec{Lorentz transformations and the time orientation}

Under a Lorentz transformations of the local frames, parametrized by an $\grpSL(2)$ valued field $\llpar$, a vector in the local frame at $x$ transforms as $\avec(x) \mapsto \llpar^{-1}(x)\avec(x)\llpar(x)$. For the dreibein and the spin connection, we have
\beq[loc-lor]
  \eee_\mu \mapsto \llpar^{-1} \eee_\mu \llpar , \qquad
  \omg_\mu \mapsto \llpar^{-1} (\del_\mu + \omg_\mu) \, \llpar.
\eeq
The metric is invariant, and the transport operator transforms as 
\beq
    \chol(x,y) \mapsto \llpar^{-1}(x) \, \chol(x,y) \, \llpar(y) .
\eeq
Such transformations will be considered as gauge symmetries. They are however not the most general transformations that leave the metric $g_{\mu\nu}$ invariant. A priori, the invariance group of the metric is $\grpO(1,2)$. Due to the determinant condition, the symmetry is reduced to $\grpSO(1,2)$. However, \eref{loc-lor} only covers its connected component $\grpSO_+(1,2)$ consisting of time oriented proper Lorentz transformations, of which the spinor representation $\grpSL(2)$ is a two-fold covering.

If we only consider this as the gauge group, then the dreibein contains more physical information than the metric. This extra ``bit'' of information has to do with the time orientation of spacetime. We can use the dreibein to distinguish between the future and the past of a point in $\sptm$. A timelike or lightlike vector $\avecv^\mu$ points towards the future, if its image in the local frame $\avec=\avecv^\mu\eee_\mu$ is causal, that is, positive timelike or lightlike. This defines a global time orientation of spacetime, which is invariant under gauge transformations of the type \eref{loc-lor}.

The definition of the time orientation of spacetime by the dreibein will be quite useful in the ADM formalism, which we are going to use for the Hamiltonian formulation and quantization of our model. Spacetime is then decomposed into $\sptm=\Rset\times\spce$, where $\spce$ is some two dimensional space manifold, and $t\in\Rset$ is a time coordinate. This time coordinate should point into the same direction as the physical time. If the surface $\spce$ is spacelike, and the dreibein determinant is positive, then this is equivalent to the statement that the normal vector of the ADM surface, which is given by $\eps^{t\mu\nu}\eee_\mu\eee_\nu$, is negative timelike. We shall refer to this as the ``ADM condition''.

\subsec{Einstein equations}

The Einstein equations in the dreibein formalism can most conveniently be derived from the first order, Palatini type Einstein Hilbert action
\beq[eh-act]
   \frac{1}{16\pi\newton} \inti{\sptm} \d^3x \, \eps^{\mu\nu\rho}\,
     \Trr{\eee_\mu \flds_{\nu\rho} }.
\eeq
The dreibein $\eee_\mu$ has the dimension of length, and the field strength $\flds_{\mu\nu}$ is dimensionless, so Newtons constant $\newton$ must have the dimension of an inverse energy, or inverse mass. This explains why there is a natural mass scale in three dimensional gravity. 

The vacuum Einstein equations derived from this action, and these are in fact the only ones we need, are the ``torsion equation'' stating that the dreibein is covariantly constant, and the ``curvature equation'' requiring the field strength of the spin connection to vanish,
\beq[eom]
  \eps^{\mu\nu\rho} \covD_\nu \eee_\rho = 0, \, \qquad
  \eps^{\mu\nu\rho} \flds_{\nu\rho} = 0.
\eeq
Together they imply that the metric is flat, which means that there are no local gravitational degrees of freedom in the theory. In particular, gravitational waves cannot exist, and there is no gravitational force. Nevertheless, if we insert matter into the universe, there will be effects that very much look like gravitational attraction or scattering.

\subsec{Embeddings}

The general solution to the vacuum Einstein equations is well known \cite{Witten,Matschull2,Grignani}. Within a simply connected region $\sbst\subset\sptm$, it is parametrized by two scalar fields $\ggg$ and $\fff$, taking values in the group $\grpSL(2)$ and the algebra $\algsl(2)$, respectively,
\beq[gen-sol]
  \omg_\mu = \ggg^{-1} \del_\mu \ggg, \qquad
  \eee_\mu = \ggg^{-1} \del_\mu \fff \, \ggg.
\eeq
The metric and the dreibein determinant in $\sbst$ becomes
\beq[metric-emb]
   g_{\mu\nu} = \case12 \Trr{\del_\mu\fff\,\del_\nu\fff}, \qquad
   \dete =  \case1{12} \eps^{\mu\nu\rho} 
            \Trr{\del_\mu\fff\, \del_\nu\fff\, \del_\rho\fff},
\eeq
and the transport operator for a curve that entirely lies inside $\sbst$ is given 
by 
\beq[hol-emb]
  \chol(x,y) = \ggg^{-1}(x) \, \ggg(y).
\eeq
From these relations we infer the following physical interpretation for $\ggg$ and $\fff$. The map $\fff$ provides an isometric embedding of $\sbst$ into Minkowski space. The metric on $\sbst$ is equal to the metric induced by the embedding, and $\fff$ is locally one-to-one, provided that the determinant is positive, which is equal to the Jacobian of $\fff$. Hence, every simply connected region of spacetime looks like a piece of Minkowski space.

There is also an interpretation for $\ggg$. Consider the expression for the transport operator above. The parallel transport is independent of the path, and it is carried out in two steps. A vector (or spinor) in the local frame at $x$ is first mapped into a ``background frame'' by $\ggg^{-1}(x)$, and from this background frame it is then mapped into the local frame at $y$ by $\ggg(y)$. The field $\ggg$ identifies all the local frames inside $\sbst$ with a single background frame, which is the one associated with the embedding Minkowski space.

So, we can say that the pair $(\ggg,\fff)$ provides an embedding of the complete bundle of local frames over $\sbst$ into Minkowski space. Every quantity, for example a vector $\avec(x)$ in the local frame at $x$, can be equivalently represented in the background frame as $\bvec=\ggg(x)\avec(x)\ggg^{-1}(x)$. To see that this is consistent, consider a local Lorentz transformation \eref{loc-lor}. Applying this to \eref{gen-sol}, we find that
\beq
  \ggg \mapsto \ggg \, \llpar ,\qquad
  \fff \mapsto \fff.
\eeq
Hence, the embedding of spacetime into Minkowski space is not affected. The vector $\bvec$ in the background frame is also unchanged, because under a local Lorentz transformation we have $\avec(x)\mapsto\llpar^{-1}(x)\avec(x)\llpar(x)$. The background space is invariant under local Lorentz transformations. But there is another symmetry group that naturally acts on it. The right hand sides of \eref{gen-sol} are invariant under rigid Poincar\'e transformations,
\beq[emb-pnc]
  \ggg \mapsto \rlpar^{-1} \ggg, \qquad
  \fff \mapsto \rlpar^{-1} ( \fff - \rtpar ) \, \rlpar,
\eeq
where $\rlpar\in\grpSL(2)$ and $\rtpar\in\algsl(2)$ are \emph{constants}. An isometric embedding of a piece of flat spacetime into Minkowski space is only determined up to such a transformation. If we think of the background space as a rest frame of some, say, external observer, then it is the location and the orientation of this observer that is changed by \eref{emb-pnc}.

The freedom to perform a rigid Poincar\'e transformation shows that the background frame, or the observer associated with it, has no particular physical meaning. It is just an auxiliary construction, which was used to write down the solutions to the vacuum Einstein equations. If we parametrize these solutions by $\ggg$ and $\fff$, then the Poincar\'e transformations \eref{emb-pnc} should be considered as gauge symmetries. This will change when we derive the action for the particle model in \sref{hamilton}. In order to obtain a well defined action, we have at add an observer to the universe, which will essentially be described by such an embedding, and this observer will become a physical reality. That is, the action will explicitly depend on the fields $\ggg$ and $\fff$, so that the background frame is no longer a gauge degree of freedom.

\section{The particle}
\label{particle}

Let us now turn to the description of a point particle. In three dimensional Einstein gravity, spacetime in the neighbourhood of a massive, pointlike particle has the shape of a cone, with the particle sitting at the top. As already mentioned in the introduction, the deficit angle $\defangle$ of the cone is related to the mass $\mass$ of the particle by $\defangle = 8\pi\newton\,\mass$. To obtain a dimensionless parameter $\mass$, and to avoid unnecessary factors in the formulas, we set $\newton = 1/4\pi$ in the following, so that $\defangle=2\mass$. In the neighbourhood of the world line of such a particle, we can introduce cylindrical coordinates $(t,r,\p)$, such that the metric becomes
\beq[rst-ds]
  \d s^2 =  - \d t^2 + \d r^2 
        + \Big( 1 - \frac{\mass}{\pi} \Big)^2 \, r^2 \, \d \p^2.
\eeq
Here, we have $r\ge0$, and the angular coordinate $\p$ is periodic with a period of $2\pi$. The total angle of this conical singularity is $2\pi-2\mass$, which is the ratio of the circumference and the radius of a small circle centered at $r=0$. So, the deficit angle is indeed equal to $2\mass$.

There are some special values for the mass which restrict the ``physically reasonable'' range of the parameter $\mass$. First of all, the deficit angle of a cone cannot be larger than $2\pi$, and thus the mass should not exceed $\pi$. In fact, we can restrict the parameter $\mass$ in \eref{rst-ds} to be less than or equal to $\pi$, without loss of generality, because replacing $\mass$ with $2\pi-\mass$ does not affect the metric. On the other hand, the mass of the particle should also be non-negative. The allowed range for the parameter $\mass$ is therefore $0\le\mass\le\pi$.

To see that negative masses lead to ``unphysical'' situations, consider the metric \eref{rst-ds} for $\mass<0$. In this case, the total angle of the conical singularity is larger than $2\pi$. This implies that there is negative curvature, and thus a negative energy density at the top of the cone. As a consequence, the particle is gravitationally repulsive. Two parallel geodesics, interpreted as test particles, passing the particle on opposite sides, are bent away from each other after they have passed the singularity. In other words, they are repelled from the matter source.

Another question is whether we should include the bounds $0$ and $\pi$ into the range of $\mass$ or not. If we consider the metric given above, then both cases are somewhat problematic. For $\mass=0$, we recover flat Minkowski space, and the particle disappears completely. For $\mass=\pi$, the metric becomes singular and the cone degenerates to a line. However, in both cases, this is only due to the special coordinate system. It describes the particle in its own rest frame. We should expect that, for $\mass=0$, the world line of the particle becomes lightlike, so that a rest frame no longer exist. This will in fact be the case, and, quite surprisingly, the same will happen at the upper bound $\mass=\pi$.

\subsec{Regularization}

For a proper description of a pointlike particle within the framework of matter coupled Einstein gravity as a field theory, we somehow have to get rid of the curvature singularity at $r=0$. There are basically two known ways to deal with this problem. The first possibility is to introduce a \emph{puncture}. That is, one excludes the world line of the particle from the spacetime manifold. The metric is then regular everywhere on the remaining manifold, and the particle is described by an appropriate set of boundary conditions imposed on the fields in the neighbourhood of the puncture. The second possibility is to \emph{regularize} the matter source, for example by replacing the pointlike particle by some, sufficiently smooth, cylindrical matter source centered around the world line. Then one does not need to impose boundary conditions, but instead one has to deal with more complicated, matter coupled Einstein equations, and in the end one has to take a limit where the radius of the cylinder shrinks to zero.

Here we shall use a new method which is somehow a mixture of these, and which makes as much as possible use of the advantages of both methods. It has in common with the puncture method that it essentially converts the matter degrees of freedom carried by the particle into topological degrees of freedom of the gravitational field, associated with a non-contractible loop that surrounds the puncture. On the other hand, it also works somewhat like the second method. The world line of the particle is smeared out, in a way which is sufficient to remove the singularity. It is however not necessary to take any kind of limit to remove the regularization. In particular, the metric in the neighbourhood of the world line will not be affected by the regularization.

The basic idea is quite simple. Consider the covariant metric given by \eref{rst-ds}. As it stands, it is not immediately obvious that there is a singularity. It is given as a smooth function of the coordinates $(t,r,\p)$ and their differentials. The \emph{curvature} singularity cannot be seen because there is also a \emph{coordinate} singularity at $r=0$. To get rid of the curvature singularity, it might be sufficient to remove the coordinate singularity. This can be done, by changing the topology of spacetime. Instead of interpreting the cylindrical coordinates in the usual way, we do not identify the points with different $\p$ at $r=0$. As a result, spacetime is no longer an $\Rset^3$, but a manifold with a \emph{cylindrical boundary}. This boundary represents the world line of the particle, which is thus, in a sense, smeared out. But there is also a non-contractible loop surrounding the particle, so that topological degrees of freedom of the gravitational field can arise.

On the modified spacetime manifold, it is now rather obvious that the metric becomes a smooth second rank tensor. The coordinate system $(t,r,\p)$ is everywhere regular. The pullback of the metric on the boundary, where $(t,\p)$ are the standard coordinates on a cylinder, is given by $\d s^2=-\d t^2$. So, the only problem is that the metric is singular on the boundary. It is of rank one instead of two, which is the dimension of the boundary. This reflects the fact that the \emph{two dimensional} cylindrical boundary actually looks like a \emph{one dimensional} world line of a particle. In the dreibein formulation of gravity, exactly these kinds of singular metrics can be consistently included. A possible dreibein representation of \eref{rst-ds} is given by
\beq[rst-ee]
  \eee_t \aln= \gam_0 , \qquad
  \eee_r = \cos\p \, \gam_1 + \sin\p \, \gam_2  , \nwl
  \eee_\p \aln= \Big( 1 - \frac\mass\pi \Big) \, r \, 
             \big( \cos\p \, \gam_2 - \sin \p \, \gam_1 \big).
\eeq
It is chosen such that the determinant is positive, the ADM condition holds (with $\eps^{tr\p}=1$), and the spin connection takes the simplest possible form,
\beq[rst-om]
  \omg_t = 0 , \qquad
  \omg_r = 0 , \qquad
  \omg_\p =  - \frac\mass{2\pi} \, \gam_0 .
\eeq
Obviously, all these fields are smooth, and they provide a solution to the \emph{vacuum} Einstein equations on the modified spacetime manifold. The curvature singularity is completely removed, although the (smeared) world line of the particle is still included in the spacetime manifold. 

We can summarize this as follows. A suitable way to include point particles into the dreibein formulation of gravity is to consider their world lines as cylindrical boundaries of spacetime. As we are, within this article, only interested in a universe containing exactly one such particle, we can fix the spacetime manifold to be $\sptm = \Rset \times \Rset_+ \times \sphr^1$. The cylindrical coordinates $(t,r,\p)$ then form a global coordinate system on $\sptm$. It is also easy to see that $\sptm$ satisfies the assumption made in the previous section, namely, that its tangent bundle is trivial. The only transition function, arising due to the identification $\p\equiv\p+2\pi$, is the identity.

The only necessary modification to the definition of dreibein gravity in the previous section is that we have to weaken the determinant and ADM conditions sightly. To allow singular metrics on the boundary, we have to require $\dete>0$ in the \emph{interior} of $\spce$ only. The metric can then be of a lower rank on the boundary, and it will in fact be forced to be of rank one by the field equations. The same also holds for the ADM condition, saying that the normal vector of the hypersurfaces of constant time coordinate should be negative timelike. For \eref{rst-ee}, this vector vanishes on the boundary as well, so that the ADM condition also applies in the interior of $\spce$ only.

\subsec{The point particle condition}

\begin{figure}[t] 
\begin{center} 
\epsfbox{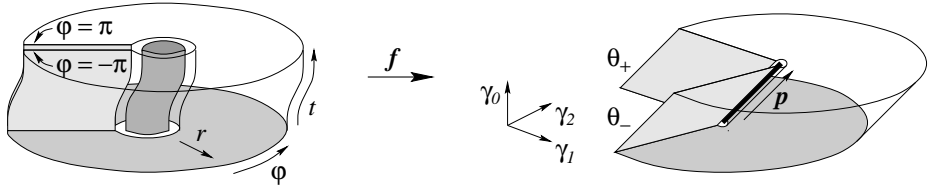} 
\caption{The embedding of a neighbourhood of the cylindrical boundary of spacetime (to the left) into the background Minkowski space (to the right). Thereby, the boundary at $r=0$ is mapped onto the world line pointing into the direction of $\mom$, the two sides of the cut at $\p=\pm\pi$ are mapped onto the surfaces $\theta_\pm$, and the ADM surface of constant $t$ is mapped onto a surface of constant $\fffv_0$.}  
\label{prt} 
\end{center} 
\hrule 
\end{figure}

The vacuum Einstein equations are not sufficient to make the boundary of $\sptm$ look like a world line of a point particle with a given mass. We need some additional field equations on the boundary. One of them has to ensure that the metric on the boundary is of rank one, so that it looks like a line. We shall refer to this as the ``point particle condition''. Let us first deal with this, and then return to the mass of the particle later on. If we want a cylinder to look like a line, then its circumference has to vanish. This means that all $\p$-components of the covariant metric have to vanish at $r=0$, or, equivalently, 
\beq[prt-con]
  \bar\eee_\p = 0.
\eeq
Here, the bar is used to denote the value of a field on the boundary. Obviously, the dreibein \eref{rst-ee} satisfies this condition. But what is the general solution to \eref{prt-con} and the Einstein equations?  For the vacuum Einstein equations, we know this already. Inside a simply connected region, it is given by an embedding $(\ggg,\fff)$ of spacetime into Minkowski space. Now, the neighbourhood of the particle is not simply connected, but we can cover it by a simply connected region. If we introduce a cut along, say, the plane at $\p=\pm\pi$ (see \fref{prt}), then we can write the general solution to the Einstein equations in the neighbourhood of the boundary as in \eref{gen-sol},
\beq[prt-emb]
  \omg_\mu = \ggg^{-1} \del_\mu \ggg , \qquad
  \eee_\mu = \ggg^{-1} \del_\mu \fff \, \ggg,
\eeq
where $\ggg$ and $\fff$ are defined on the range $-\pi\le\p\le\pi$, and, of course, they also depend on $t$ and $r$. Their values at $\p=\pm\pi$, denoted by $\ggg_\pm$ and $\fff_\pm$, are not independent. They need not be equal, but the one-forms $\eee_\mu$ and $\omg_\mu$ must be continuous, which means that the right hand sides of \eref{prt-emb} at $\p=-\pi$ must coincide with those at $\p=\pi$. We saw that this is the case if and only if the fields $\ggg$ and $\fff$ are related by a Poincar\'e transformation \eref{emb-pnc}. Hence, we must have
\beq[prt-emb-cut]
  \ggg_+ = \hol^{-1}  \ggg_- , \qquad
  \fff_+ = \hol^{-1}  (\fff_- - \ang) \, \hol,
\eeq
where $\hol\in\grpSL(2)$ and $\ang\in\algsl(2)$ are constants. In addition to this, we also have to take into account the point particle condition. Expressed in term of $\ggg$ and $\fff$, it becomes a simple differential equation,
\beq[prt-con-emb]
  \bar \eee_\p = \bar\ggg^{-1} \del_\p \bar\fff \, \bar\ggg = 0 ,
\eeq
which states that the value of $\fff$ on the boundary must be a function of $t$ only, $\bar\fff(t,\p) = \xpos(t)$. This function is further restricted by the condition \eref{prt-emb-cut}, which yields
\beq[prt-ang]
  \ang = \xpos(t) - \hol \, \xpos(t) \, \hol^{-1} ,
\eeq
because $\bar\fff_+(t)=\bar\fff_-(t)=\xpos(t)$. Now, $\hol$ and $\ang$ are constants, and therefore the time derivative of this equation implies that the time derivative of $\xpos$ must commute with $\hol$. All vectors that commute with a given group element $\hol\in\grpSL(2)$ are proportional to its ``projection'' $\mom\in\algsl(2)$ into the algebra (see \eref{projection}). This vector $\mom$ is also the ``axis'' of the Lorentz transformation represented by $\hol$. It can be obtained by expanding the matrix $\hol$ in terms of the unit and gamma matrices, and then dropping the contribution proportional to the unit matrix,
\beq[prt-hol-mom]
    \hol = \moms \, \one + \momv_a \, \gam^a 
    \quad \mapsto \quad
    \mom = \momv_a \gam^a .
\eeq
Taking all this together, we find that the most general function $\fff$ is, on the boundary, of the form
\beq[prt-emb-bnd]  
  \bar\fff(t,\p) = \ypos + \tau(t)\, \mom,
\eeq
where $\ypos\in\algsl(2)$ is some constant and $\tau$ is an arbitrary real function of time. Obviously, this reminds of the description of a relativistic point particle in Minkowski space. The particle has a momentum $\mom$, and it passes through the point $\ypos$. Indeed, with the interpretation of $\fff$ as an embedding of spacetime into Minkowski space, $\bar\fff$ is the position of the particle in this background space. If it is considered as the rest frame of an external observer, then he sees the particle moving along the world line \eref{prt-emb-bnd}. The arbitrary function $\tau(t)$ has physical interpretation as well. Consider the time component of the dreibein on the boundary,
\beq[bnd-ee-t]
  \bar\eee_t = \bar\ggg^{-1} \del_t \bar\fff \, \bar\ggg = 
              \del_t \tau \, \bar\ggg^{-1} \mom \, \bar\ggg .
\eeq
It represents the tangent vector of the world line, or the velocity of the particle, in the local frames on the boundary. Its length is given by
\beq
    \sqrt{- \bar g_{tt}}  
  = | \del_t \tau | \, \sqrt{-\momv_a\momv^a} ,
\eeq
and this is the quotient of the physical time elapsing on the world line divided by the coordinate time. The function $\tau$ is thus, up to a factor, the eigentime of the particle.

\subsec{The mass shell condition}

To see that the vector $\mom$, or rather the group element $\hol$ from which it is derived, behaves very much like the momentum of a relativistic point particle, let us now, before we come back to the solution of the field equations in the neighbourhood of the particle, consider the second boundary condition to be imposed. So far, we haven't fixed the mass of the particle. It should be related to the deficit angle by $\defangle=2\mass$, and, like the point particle condition, we want to write this ``mass shell condition'' as a local field equation, that is, it should involve the fields on the boundary only.

At first sight, it is not so obvious how this can be achieved, because the deficit angle, defined as the missing circumference of a small circle, is a property of the metric \emph{in the neighbourhood} of the particle. However, if we slightly change the definition of the mass, we can express it as a function of the spin connection \emph{on the boundary}. The idea is that, instead of measuring the radius and circumference of a circle, we can also perform a parallel transport of a vector (or spinor) around the particle. If the deficit angle is $2\mass$, then the result should be that the vector is rotated by $2\mass$.

The parallel transport along a given curve is described by the transport operator \eref{hol-def}. Provided that the curvature vanishes, this is invariant under smooth deformations of the curve. If we are interested in the effect of transporting a vector or spinor around the particle, we can, in particular, perform this transport on the boundary. Let us therefore define an $\grpSL(2)$ valued field $\bar\hol(t,\p)$, which describes the transport of a spinor once around the boundary, on a circle of constant $t$, starting and ending at the point $\p$. For later convenience, we choose the ``clockwise'', that is, negative $\p$ direction. The ``holonomy'' of the particle is then given by
\beq[prt-hol]
   \bar\hol^{-1}(t,\p) = 
    \Pexp \intl{\p}{\p+2\pi} \d \tilde\p \,\, \bar\omg_\p(t,\tilde\p).
\eeq
Independently of whether the spin connection is flat or not, this is a well defined field on the boundary. It is, by definition, covariantly constant with respect to $\p$,
\beq[prt-hol-con]
   \covD_\p \bar\hol = \del_\p \bar\hol + [\bar\omg_\p,\bar\hol] = 0,
\eeq
but there is, a priori, no relation between the values of $\bar\hol$ at different times $t$. We can therefore think of $\bar\hol$ as a function on the world line of the particle, which ``essentially'' depends on $t$ only. The dependence of $\bar\hol$ on $\p$ is only due to the fact that the same physical quantity is represented in different local frames on the boundary.

If the Einstein equations are satisfied, then $\bar\hol$ is covariantly constant with respect to $t$ as well. The physical quantity represented by $\bar\hol$ is thus a constant of motion. It is in fact the same as the constant $\hol$ introduced above, which we interpreted as the \emph{momentum} of the particle. The relation between $\bar\hol$ and $\hol$, or that between their projections $\mom$ and $\bar\mom$, is found to be 
\beq[prt-hol-bnd]
   \bar\hol = \bar\ggg^{-1} \hol \, \bar\ggg, \qquad
   \bar\mom = \bar\ggg^{-1} \mom \, \bar\ggg.
\eeq
The right hand sides are continuous at $\p=\pm\pi$, as a consequence of \eref{prt-emb-cut}, although $\bar\ggg$ is not. These relations are also the reason for choosing the clockwise direction in the definition of $\bar\hol$. Otherwise, there would be an ``inverse'' showing up in \eref{prt-hol-bnd}.

Now, coming back to the mass shell condition, we have to specify all holonomies $\bar\hol$ which represent rotations by $2\mass$. This is a simple geometric exercise, which is explicitly carried out in \ref{lorentz}. The result is that all such elements of $\grpSL(2)$ can be characterized by the equation \eref{rot-con},
\beq[mss-con]
    \case12 \Trr{\bar\hol} = \cos\mass,
\eeq
The solutions to this equation lie in the conjugacy classes of the group elements $\expt{\pm\mass\gam_0}$. Geometrically, the sign in the exponent distinguishes between rotations in positive and negative direction. In the context of particle physics, there is also another interpretation of the sign ambiguity. Let us write this ``mass shell condition'' as a condition to be imposed on the momentum vector $\mom$. We first notice that the momentum $\hol$ in the background frame satisfies the same mass shell condition as $\bar\hol=\bar\ggg^{-1}\hol\,\bar\ggg$, because the conjugation with $\bar\ggg$ drops out under the trace. Furthermore, if we use the expansion \eref{prt-hol-mom}, and that the vector $\momv_a$ is related to the scalar $\moms$ by
\beq[hyp-con]
          \moms^2 = \momv_a\momv^a + 1,
\eeq
then we find 
\beq[mss-con-vec]
  \moms = \cos\mass \follows
  \momv_a\momv^a + \sin^2\mass = 0 .
\eeq
This looks like the mass shell condition of a relativistic point particle, except that the mass is replaced by its sine. This is, in the first place, a matter of convention, and follows from the fact that we defined the momentum vector $\mom$ to be the \emph{projection} of $\hol$ into the algebra. One can also define it such that $\hol$ is the \emph{exponential} of $\mom$ \cite{Carlip2}. In this case, there would be no sine in \eref{mss-con-vec}, but we would run into trouble later on, because the exponential map is neither one-to-one, nor does it map the algebra \emph{onto} the group, and so $\mom$ would not be a well defined phase space function. In any case, the structure of the solutions to \eref{mss-con-vec} is always the same. They split into two subsets, the ``particle'' and the ``antiparticle'' mass shell, consisting of the positive and negative timelike vectors of length $\sin\mass$.

\begin{figure}[t]
\begin{center} 
\epsfbox{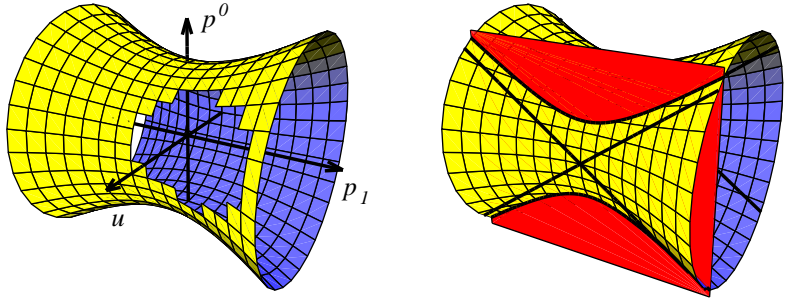} 
\caption{The group manifold $\grpSL(2)$, embedded in $\Rset^4$, using the coordinates $\momv_A=(\momv_a,\moms)$, with $\momv_2$ suppressed. The picture to the right shows the two mass shells for $\mass=\pi/6$ (a deficit angle of $60^\circ$), obtained by intersecting the group manifold with the plane $\moms=\cos\mass$, and the light cones emerging from $\one$ and $-\one$ (on the back). The grid lines on the group manifold are the Euler angles $\eurh$ and $\euch$.}
\label{grp} 
\end{center} 
\hrule 
\end{figure}

To visualize these mass shells, let us use the coordinates $(\momv_a,\moms)$ to embed the group manifold into $\Rset^4$. The condition \eref{hyp-con} defines a hyperboloid therein, which is shown in \fref{grp}. The mass shells are the intersections of this hyperboloid with the plane $\moms=\cos\mass$. We see that there is an upper and a lower mass shell, corresponding the positive and negative timelike vectors $\mom$, and that they look very similar to those of a relativistic point particle. But there are also some features that are different.

One essential difference is that the range of $\mass$ is bounded from below \emph{and} from above. We can now see that this is because the momentum $\hol$ lives on the group manifold $\grpSL(2)$, and not in flat Minkowski space. If the mass approaches the lower bound $\mass=0$, then the mass shells approach the ``light cones'' emerging from $\hol=\one$. They consist of those elements of the group for which $\moms=1$, so that $\mom$ is a lightlike vector. But now there is a second pair of light cones, emerging from the group element $\hol=-\one$. There, we have $\moms=-1$, and so $\mom$ is also lightlike. For $\mass=\pi$, the mass shells coincide with these light cones.

Between these two pairs of light cones, there is only a finite range of $\moms$, which coincides with the range of the cosine in the mass shell condition. The whole range of timelike momenta $\hol$ is covered by $0<\mass<\pi$, and on both sides of this interval the momentum becomes lightlike. Using the conventional terminology, we can say that the particle is ``massless'' for $\mass=0$ as well as for $\mass=\pi$. This is what we already mentioned in the beginning. It is the reason why the description of the particle in its own rest frame fails for these values of $\mass$.

To see what goes wrong if we take the limit $\mass\to0$ or $\mass\to\pi$ in the rest frame of the particle, consider the dreibein \eref{rst-ee} and the spin connection \eref{rst-om}. The momentum is then given by
\beq[rst-hol]
  \bar\hol = \expo{\mass\gam_0} = 
   \cos\mass\,\, \one + \sin\mass \,\, \gam_0. 
\eeq
In the limits $\mass\to0$ and $\mass\to\pi$, we have $\bar\hol\to\pm\one$ and therefore $\bar\mom\to0$. The same happens for a relativistic point particle if we take the limit $\mass\to0$ in the rest frame. The momentum does not end up on the light cone, but vanishes. To get a proper description of lightlike particles, we have to exclude the special solutions $\bar\hol=\pm\one$ of the mass shell condition. The neighbourhood of a lightlike particle will then, as we shall see in a moment, not be very different from that of a massive particle.

\subsec{The neighbourhood of the particle}

Let us now return to the general solution of the Einstein equations in the neighbourhood of the particle. So far, we have solved the point particle condition and the Einstein equations \emph{on the boundary}. To see what spacetime looks like in the neighbourhood, we have to find an embedding $(\ggg,\fff)$, such that it satisfies the condition \eref{prt-emb-cut} along the cut at $\p=\pm\pi$,
\beq[prt-emb-cut-2]
  \ggg_+ = \hol^{-1}  \ggg_- , \qquad
  \fff_+ = \hol^{-1}  (\fff_- - \ang) \, \hol ,
\eeq
and such that, on the boundary, $\fff$ takes the form \eref{prt-emb-bnd}, 
\beq[prt-emb-bnd-2]
   \bar\fff(t,\p) = \ypos + \tau(t) \, \mom . 
\eeq
Moreover, it is subject to some additional conditions. The resulting metric must be invertible, and the surfaces of constant $t$ must be spacelike. We can further restrict it by imposing ``gauge conditions''. In particular, we are still free to choose the coordinates $(t,r,\p)$ off the boundary as we like, and we can use this to make the function $\fff$ look as simple as possible.

We found that the determinant condition for the dreibein is equivalent to the statement that $\fff$ is locally one-to-one. We can therefore use $\fff$ to \emph{define} the coordinates. Let us fix the time coordinate $t$ such that the surfaces of constant $t$ in $\sptm$ are mapped onto equal time planes in the background Minkowski space. By this, we mean the planes of constant $\fffv_0=\case12 \Trr{\fff\gam_0}$. The value of the zero component of $\fff$ is then determined by its value on the boundary.

Furthermore, we can choose the radial coordinate $r$ to be the physical distance from the particle. The difference of $\fff$ to its value on the boundary is then a spacelike vector of length $r$, lying in the $\gam_1$-$\gam_2$ plane. The only remaining freedom is the direction of this vector, which can be specified by a real function $\theta(t,r,\p)$. All together, the most general function $\fff$ with all these properties is of the form
\beq[ff-off]
   \fff(t,r,\p) = \ypos + \tau \, \mom +  
          r \, \expo{ \frac12\theta \,\gam_0} \, \gam_1 
               \expo{-\frac12\theta \,\gam_0} ,
\eeq
or, in components, 
\beq
      \fffv_0 = \yposv_0 + \tau \, \momv_0 , \qquad
          \fffv_1 \aln= \yposv_1 + \tau \, \momv_1  + r \cos \theta , \nwl
          \fffv_2 \aln= \yposv_2 + \tau \, \momv_2  + r \sin \theta .
\eeq
The situation is sketched in \fref{prt}. The neighbourhood of the boundary, which is cut along the surface at $\p=\pm\pi$, is mapped onto a ``wedge'' in Minkowski space. Thereby, the ADM surface shown on the left is mapped onto the equal time surface on the right. The images of the cut at $\p=\pm\pi$ are marked $\theta_\pm$. They will turn out to be planes, as a consequence of the boundary conditions for $\fff$, but for the moment they are arbitrary surfaces described by two functions,
\beq
  \theta_+(t,r) = \theta(t,r,\pi), \qquad
  \theta_-(t,r) = \theta(t,r,-\pi). 
\eeq
Let us now check the ADM and the determinant condition. The normal vector of the constant time surface is found to be
\beq
    \case12 \eps^{t\mu\nu} \eee_\mu \eee_\nu = 
        \eee_{[r} \eee_{\p]} = 
        \ggg^{-1}  \del_{[r} \fff \, \del_{\p]}\fff  \, \ggg = 
        - r \, \del_\p \theta \, \ggg^{-1} \gam_0 \ggg,      
\eeq
and the dreibein determinant is 
\beq
   \dete = \case12 \Trr{\eee_t \, \eee_r \, \eee_\p } =
           \case12 \Trr{\del_t\fff\,\del_r\fff\,\del_\p\fff} = 
                r \, \momv^0 \,  \del_t \tau \, \del_\p \theta  .
\eeq
For $r>0$, the normal vector must be negative timelike and the determinant has to be positive. This is the case if and only if
\beq[prt-adm-det]
  \del_\p \theta > 0, \qquad
  \momv^0 \, \del_t \tau > 0.
\eeq
The first inequality says that $\theta$ must be an increasing function of $\p$, which means that the embedding shown in \fref{prt} is such that, with increasing $\p$, we move counter-clockwise around the world line of the particle in the embedding space. In particular, we must have $\theta_+>\theta_-$. The second one implies that $\tau$ is an increasing or decreasing function of time, depending on whether $\momv^0$ is positive or negative. It cannot be fulfilled if $\momv^0$ is zero, but then the momentum vector would be either spacelike or zero, which is excluded by the mass shell condition. 

The relation between $t$, $\tau$ and the sign of $\momv^0$ confirms very nicely the interpretation of the states with positive and negative $\momv^0$ as particle and antiparticle states. Particles have positive timelike or lightlike momentum vectors, and their eigentime $\tau$ is increasing with the external time $t$. For antiparticles, the momentum vector is negative timelike or lightlike, and the eigentime is running backwards. This is another analogy to the relativistic point particle.

Now, we still have to check the boundary condition \eref{prt-emb-cut-2}. For this purpose, it is most convenient to express the holonomy $\hol\in\grpSL(2)$ in terms of ``Euler angles'' (see \eref{euler}). They and provide a global set of coordinates on the group manifold $\grpSL(2)$. A generic group element can be written as
\beq[prt-hol-euler]
  \hol     \aln=    \expo{\frac12 (\eurh+\euph)\gam_0} \, 
                    \expo{         \euch \gam_1} \, 
                    \expo{\frac12 (\eurh-\euph)\gam_0} \nwl 
      \aln=  \cosh\euch \, ( \cos\eurh \,\, \one   
                              +  \sin\eurh \,\, \gam_0 )
           + \sinh\euch \, ( \cos\euph \,\, \gam_1 
                              +  \sin\euph \,\, \gam_2 ).
\eeq
Here, $\eurh$ and $\euph$ are periodic ``angular'' coordinates with period $2\pi$, and $\euch$ is a positive ``radial'' coordinate, which forms together with $\euph$ a polar coordinate system (for $\euch=0$ the angle $\euph$ is redundant). The coordinates $\eurh$ and $\euch$ are shown in \fref{grp}, as the grid lines. The angular coordinate $\eurh$ is the one that winds around the hyperboloid, and $\euch$ increases with the distance from the center. We can also express the momentum vector $\mom$ in terms of these variables. It becomes
\beq[prt-mom-euler]
  \mom = \cosh\euch \sin\eurh \,\, \gam_0 
         + \sinh\euch \, 
              ( \cos \euph \,\, \gam_1 + \sin\euph \,\, \gam_2 ),
\eeq
and this tells us that the particle is moving with a velocity of ${\tanh\euch}/{\sin\eurh}$ (which is $1$ for lightlike and smaller than one for timelike vectors $\mom$) into the spatial direction specified by the angle $\euph$. We can distinguish particles and antiparticles using the coordinate $\eurh$. We have a particle state for $0<\eurh<\pi$, and an antiparticle state for $-\pi<\eurh<0$. As $\eurh$ has a period of $2\pi$ and cannot be a multiple of $\pi$ (in this case $\momv^0$ would be zero), it always lies in one of these intervals.

Inserting all this into the boundary condition \eref{prt-emb-cut-2} for $\fff$, we obtain a relation between the functions $\theta_\pm$ and the Euler angles, which after some algebraic manipulations becomes
\beq[prt-cut-con]
    \expo{- \frac12 (\theta_++\eurh+\euph)\gam_0}  
    \expo{- \euch\gam_1}  
    \expo{  \frac12 (\theta_--\eurh+\euph)\gam_0}  
    \gam_1 
    \expo{- \frac12 (\theta_--\eurh-\euph)\gam_0} 
    \expo{  \euch\gam_1}  
    \expo{  \frac12 (\theta_++\eurh-\euph)\gam_0}  = \gam_1. \quad
\eeq
Let us first consider the case where $\euch=0$, describing a particle which is at rest in the embedding Minkowski space. In this case, \eref{prt-cut-con} reduces to
\beq
    \expo{-\frac12 (\theta_+ - \theta_- + 2 \eurh ) \gam_0} \, 
    \gam_1 \, 
    \expo{ \frac12 (\theta_+ - \theta_- + 2 \eurh ) \gam_0}  = \gam_1 .
\eeq
According to the formula \eref{exp-rot}, this implies that the term between the parenthesis must be a multiple of $2\pi$, so that
\beq[tot-ang]
    \theta_+ - \theta_- = 2\pi z - 2\eurh, \qquad z \in \Zset.
\eeq
In particular, the difference $\theta_+ - \theta_-$ must be a constant. This is the total angle of the cone surrounding the particle. It is determined by the momentum of the particle up to a multiple of $2\pi$. The same holds for the deficit angle, which is given by $\defangle = 2\pi - \theta_+ + \theta_-$. The integer $z$ is however not completely arbitrary. For $\theta_+ > \theta_-$, we must have $z\ge1$ for particles and $z\ge0$ for antiparticles. All together, we find that the possible values for the deficit angle are given by
\beq[def-ang-part]
  \defangle =  2 \eurh , 
        \quad  2 \eurh - 2 \pi, 
        \quad  2 \eurh - 4 \pi, \quad \dots
\eeq
for a particle state with $0<\eurh<\pi$, and 
\beq[def-ang-anti]
  \defangle =  2 \eurh + 2 \pi , 
        \quad  2 \eurh , 
        \quad  2 \eurh - 2 \pi , \quad \dots
\eeq
for an antiparticle state with $-\pi<\eurh<0$. Obviously, only the first elements of these series are positive. In all other cases, the deficit angle is negative. The total angle is then bigger that $2\pi$, which means that the wedge shown in \fref{prt} overlaps itself. In the beginning, we argued that such particles are ``unphysical'', because they are gravitationally repulsive. To exclude them, we can impose yet another boundary condition on the metric. 

It is not necessary to specify this condition explicitly as, say, a function of the metric in the neighbourhood of the particle. It is sufficient to know that, for any given momentum $\hol$, there is a discrete series of solutions to the Einstein equations in the neighbourhood of the particle, and that the condition $\defangle>0$ picks out exactly one of these solutions. To be precise, so far we have only shown this for a particle at rest, with $\euch=0$. For $\euch>0$, the condition \eref{prt-cut-con} is more restrictive. Evaluating it using the formulas \eref{exp-gam}, one finds that
\beq[prt-cut-dir]
  \theta_+ = \euph - \eurh + 2\pi z_+, \qquad
  \theta_- = \euph + \eurh - 2\pi z_-,
\eeq
where $z_\pm$ are either two integers or two half integers. In particular, the right hand sides of these equations are constants, and thus the surfaces $\theta_\pm$ must be planes. Subtracting the two equations, we get \eref{tot-ang} back. The relation between $\eurh$ and the deficit angle is therefore also valid for moving particles.

The additional information contained in \eref{prt-cut-dir} is the following. It is not only the total angle that is determined by the momentum, but also the directions $\theta_+$ and $\theta_-$ themselves are fixed. If we add the two equations, then we find that the angle between the directions $\theta_+$ and $\euph$ must be the same as that between $\euph$ and $\theta_-$. As the Euler angle $\euph$ was the direction in which the particle is moving, this means that the two images of the cut in the embedding Minkowski space have to lie symmetrically ``in front of'' or ``behind'' the particle. The latter situation is shown in the figure.

This reproduces very nicely an observation made by 't~Hooft \cite{Hooft4}, who obtained the same result in a somewhat different way. If one wants the ADM surfaces to become equal time planes in an observer's rest frame, which is not the rest frame of the particle, then the wedge that the particle ``cuts out'' from Minkowski space has to be placed either straight in front of or behind the particle. We should however emphasize that this is just a result of the special gauge choice, that is, the special choice of coordinates in the neighbourhood of the particle. This was only made to show what this neighbourhood looks like. We shall not impose any such gauge fixing on our model.

\subsec{The rest mass and the energy}

We have now found a relation between the momentum $\hol$ of the particle and the deficit angle of the conical singularity on its world line. But this seems to be somewhat in contradiction with the previously considered relation $\defangle=2\mass$. Instead of this, we now have $\defangle=2\eurh$ for particles with $0<\eurh<\pi$, and $\defangle=2\pi+2\eurh$ for antiparticles with $-\pi<\eurh<0$. Furthermore, the deficit angle is \emph{frame dependent}. It depends on the relative motion of the particle with respect to the observer. To find a relation between the deficit angle and the mass, we have to express the mass shell condition in terms of the Euler angles. It becomes
\beq[mss-con-euler]
    \cosh\euch \, \cos\eurh = \cos\mass.
\eeq
It provides a relation between the mass $\mass$, the ``rapidity'' $\euch$, which measures the velocity of the particle with respect to the observer, and the deficit angle. If the particle is at rest, then $\euch$ vanishes and this reduces to $\cos\eurh=\cos\mass$. There are two solutions to this equation, $\eurh=\pm\mass$. For the positive sign, we have a particle state, and we recover the original relation $\defangle=2\mass$. If we choose the negative sign, then we have an antiparticle state, and the deficit angle becomes $\defangle=2\pi-2\mass$. 

Actually, one should expect the gravitational field of an antiparticle to be the same as that of the corresponding particle. But, in a somewhat heuristical way, it can be understood as a result of the unusual spectrum of the parameter $\mass$. Antiparticles are related to particles by a charge conjugation operation, which inverts all charges, including the gravitational charge, and thus the mass. But the only reasonable way to ``invert'' the mass, which takes values in the range $0\le\mass\le\pi$, is to replace it by $\pi-\mass$.

Apart from this somewhat strange effect, a more interesting question is, what happens if we leave the rest frame of the particle. The deficit angle is then no longer equal to $2\mass$ or $2\pi-2\mass$. Geometrically, this can be understood as follows. If the particle is accelerated, then the wedge in \fref{prt} is boosted, so that the angle between $\theta_+$ and $\theta_-$ changes. This explains why the deficit angle is a dynamical quantity. Let us see how it behaves for small masses and small velocities. We can then expand the mass shell condition \eref{mss-con-euler}, and what we get is
\beq[mss-con-exp]
   (1+\case12 \euch^2) (1 - \case12 \eurh^2) \approx (1-\case12\mass^2) 
   \follows
   \eurh^2 \approx  \euch^2 + \mass^2 .
\eeq
Now, if $\euch$ is small, then it is approximately the length of the spatial component of the momentum \eref{prt-mom-euler}, and the above relation becomes the usual relation between the mass, the spatial momentum, and the energy of a relativistic point particle. This suggests that we should interpret the deficit angle as the \emph{energy} of the particle rather than its \emph{mass}. The relation \eref{mss-con-euler} is then a somewhat deformed dispersion relation, which reduces to the usual one in the low energy limit.

At large scales, the behaviour of $\eurh$ as a function of $\euch$ is however completely different. For large $\euch$, $\eurh$ always saturates at $\pi/2$ for particles and at $-\pi/2$ for antiparticles. Whether it approaches this limit from below or from above depends on whether its rest mass is smaller or bigger than $\pi/2$. This can also be seen in \fref{grp}, where the grid lines on the hyperboloid represent the Euler angles. For large momenta, the value of $\eurh$ on the upper mass shell always approaches that on the ``ridge'' of the hyperboloid, which is $\pi/2$. As a result, the energy spectrum of the particle is always bounded from below \emph{and} from above, and it ranges from $\mass$ to $\pi/2$ for particles and from $-\mass$ to $-\pi/2$ for antiparticles.

A peculiar situation arises when $\mass=\pi/2$. The energy $\eurh$ is then independent of the velocity of the particle. In this case, the deficit angle is always equal to $\pi$, so that, referring again to \fref{prt}, exactly half of the Minkowski space is cut out. Clearly, every observer will then agree with this fact. It is also the limit that cannot be transgressed by accelerating a particle with a mass different from $\pi/2$. Hence, if the mass is smaller than $\pi/2$ (or bigger, for an antiparticle), then the deficit angle is always smaller than $\pi/2$, and vice versa.

\subsec{Lightlike particles}

Finally, let us briefly come back to the two types of lightlike particles, with $\mass=0$ and $\mass=\pi$, and let us, for simplicity, consider the particle states only. The antiparticles behave similarly, if one replaces $\mass=0$ with $\mass=\pi$ and vice versa. For a particle with $\mass=0$, the mass shell condition implies that $0<\eurh<\pi/2$, and thus the deficit angle lies in the range $0<\defangle<\pi$. This is the situation shown in \fref{prt}, if we take $\mom$ to be a lightlike vector. There is nothing special about this spacetime, except that it is not possible to transform to a rest frame of the particle. The observer simply sees a particle moving with the speed of light.

Something rather different happens for $\mass=\pi$. Then the deficit angle lies in the range $\pi<\defangle<2\pi$, and the two planes in \fref{prt} point to the right, into the direction in which the particle is moving with the speed of light. The crucial difference to the former situation is that it is now, for a test particle moving on a straight timelike line, not possible to ``escape'' the planes $\theta_\pm$. After some finite time, it will be hit by one of them, say $\theta_-$, and reappear at $\theta_+$. Then, again after some finite time, the same happens again, and so on. One can show that the intervals between these events become smaller and smaller, and that the test particle will finally collide with the lightlike particle.

Expressed differently, a particle with $\mass=\pi$ is a rather strong gravitational attractor, although there is no local gravitational force. But nevertheless, everything moving on timelike lines will hit the attractor after some finite time. It is almost something like a ``big crunch'' singularity of the universe, where every timelike and every lightlike geodesic ends on, except that lightlike geodesics can still escape the singularity. For a different way to show that this is a typical behaviour of a lightlike conical singularity, we refer to section~3.4 in \cite{Matschull2}. There, it is also shown that for spacelike holonomies $\hol$, the singularity essentially becomes a big crunch (or big bang) singularity. This is another way to see that $\mass=\pi$ is an upper bound for the mass. The conical singularity on the world line of the particle is then already something in between a matter source and a big crunch. It is the strongest possible gravitational attractor which is located on a causal world line.

\section{Hamiltonian formulation} 
\label{hamilton}

In this section, we want to derive the classical Hamiltonian formalism for the point particle, coupled to Einstein gravity as a field theory. What we need for this purpose is a proper definition of the configuration space of the system, and an action principle. The action should be given by the Einstein Hilbert action for the gravitational fields, plus an appropriate term for the particle. We shall also need some fall off conditions and boundary terms at infinity, to make the action finite and to get the correct equations of motion. Furthermore, it turns out that we have to add some additional degrees of freedom to the model, and these can most naturally be interpreted as an observer at infinity.

All this is most conveniently carried out if we switch to the ADM formulation from the very beginning. We split our spacetime manifold into $\sptm=\Rset\times\spce$, where $\spce=\Rset_+\times\sphr^1$ is the space manifold. It has a circular boundary, which represents the position of the particle at a given moment of time. On $\spce$, we introduce formal tangent indices $i,j,\dots$, taking the values $r$ and $\p$, a Levi Civita tensor $\eps^{ij}$ with $\eps^{r\p}=1$, and we denote derivatives with respect to $t$ by a dot. The one-forms $\eee_\mu$ and $\omg_\mu$ on $\sptm$ split into one-forms $\eee_i$ and $\omg_i$ on $\spce$, and scalars $\eee_t$ and $\omg_t$. The Einstein Hilbert action becomes the time integral of the Lagrangian
\beq[lag]
  \lag = \inti{\spce} \d^2x \, \eps^{ij} \, 
    \case12 \Trr{ \dot\omg_i\eee_j + 
       \omg_t \covD_i \eee_j + \case12 \eee_t \flds_{ij} }, 
\eeq
and the vacuum Einstein equations split into \emph{constraints}
\beq[constraints]
  \eps^{ij} \flds_{ij} = 0, \qquad
  \eps^{ij} \covD_i \eee_j = 0,
\eeq
and \emph{evolution equations}
\beq[evolve]
  \dot \omg_i = \covD_i \omg_t , \qquad
  \dot \eee_i = \covD_i \eee_t + [\eee_i,\omg_t] .
\eeq
In \eref{lag}, we already performed an integration by parts, such that the derivatives do not act on the time components of the fields. As we have to add boundary terms anyway, we may equally well use this as an ansatz for the total Lagrangian. It has the advantage that we can immediately read off the constraints by varying $\lag$ with respect to the ``multipliers'' $\eee_t$ and $\omg_t$. To derive the evolution equations, however, we have to take into account a boundary term. Varying the Lagrangian with respect to the ``dynamical'' fields $\eee_i$ and $\omg_i$ yields
\beq[lag-var]
  \delta\lag \aln=  \inti\spce \d^2x \, \eps^{ij} \, 
    \case12 \Trrr{  (\dot\omg_i - \covD_i \omg_t ) \, \delta \eee_j 
       + (\dot\eee_i - \covD_i \eee_t - [\eee_i,\omg_t] ) \, \delta \omg_j } 
         \nwl \aln{} {}  + 
          \inti\spce \d^2x \,  \eps^{ij} \,\case12 \,
            \del_i \Trr{ \omg_t \, \delta\eee_j + \eee_t \, \delta\omg_j }.
\eeq
To be precise, there is also a total time derivative that has been neglected, but this is of course allowed, because we are only interested in the Euler Lagrange equations. From the first line in this expression, we deduce the evolution equations. The second line can be converted into a boundary integral. There are two boundaries of $\spce$ that contribute, namely the ``particle boundary'' at $r=0$, and the ``boundary at infinity'', where $r\to\infty$. For a generic one-form $\xi_i$ on $\spce$, we have, according to Stokes' theorem,
\beq[stokes]
     \inti{\spce} \d^2x \, \eps^{ij} \, \del_i \xi_j 
     = \inti{r\to\infty} \d\p \,\, \xi_\p 
     - \inti{r=0} \d\p\,\, \bar\xi_\p. 
\eeq
Let us, for the moment, ignore the boundary at infinity, which will be considered in detail later on. The contribution to $\delta\lag$ from the particle boundary is then given by
\beq[lag-var-bnd]
     - \inti{r=0} \d \p \, \case12
     \Trr{ \bar\omg_t \, \delta \bar\eee_\p 
          + \bar\eee_t \, \delta \bar\omg_\p }.
\eeq
As this has to vanish for a generic variation of the fields, we get additional equations of motion on the boundary, namely $\bar\eee_t=0$ and $\bar\omg_t=0$. Clearly, this is not what we want. Instead, we want the point particle condition and the mass shell condition to be imposed.

\subsec{The particle action}

As an ansatz, let us introduce Lagrange multipliers $\mult$ and $\bar\xmult$, and add a constraint term to the Lagrangian
\beq[lag-prt-ans]
  \lagprt = 
   - \mult  ( \case12 \Trr{\bar\hol} - \cos\mass )
    - \inti{r=0} \d\p \, \case12 \Trr{\bar\xmult \, \bar\eee_\p}.
\eeq
Note that, to impose the point particle condition, we need an $\algsl(2)$ valued field $\bar\xmult$ on the boundary, but for the mass shell constraint it is sufficient to introduce a single real variable $\mult$. This is because the trace of the holonomy of the particle is, by definition, independent of $\p$ and thus, at a given time $t$, the mass shell condition is only a single real equation.

If we now vary $\lagprt$ with respect to the multipliers, we get the correct constraints. But we also get a contribution from $\lagprt$ to the variation with respect to the gravitational fields. To compute this, we need the derivative of the holonomy \eref{prt-hol} with respect to the spin connection, which is given by
\beq
  \case12 \, \delta \, \Trr{\bar\hol} = 
     -\inti{} \d\p \, \case12 \Trr{\bar\hol \, \delta\bar\omg_\p} 
   = -\inti{} \d\p \, \case12 \Trr{\bar\mom \, \delta\bar\omg_\p}.
\eeq
The last equality holds because $\bar\omg_\p$ is traceless, so that only the projection $\bar\mom$ of the holonomy $\bar\hol$ contributes to the trace. Using this, we find
\beq[lag-prt-var]
  \delta \lagprt =  \inti{r=0} \d\p \, 
          \case12 \Trr{\mult \, \bar\mom \, \delta \bar\omg_\p 
                           - \bar\xmult \, \delta \eee_\p}, 
\eeq
and adding this to \eref{lag-var-bnd}, the extra field equations on the boundary become
\beq[bnd-sec]
    \bar\omg_t = - \bar\xmult , \qquad
    \bar\eee_t = \mult \, \bar\mom.
\eeq
Now, the first equation only fixes the values of the multiplier field $\bar\xmult$. We can get rid of this, if we identify it from the very beginning with $\bar\omg_t$. Hence, we redefine the particle Lagrangian to be
\beq[lag-prt]
  \lagprt = 
   - \mult  ( \case12 \Trr{\bar\hol} - \cos\mass )
    + \inti{r=0} \d\p \,  \case12 \Trr{\bar\omg_t \, \bar\eee_\p}.
\eeq
We are then left with the second equation in \eref{bnd-sec}. Except for the fact that it fixes the value of the multiplier $\mult$, this turns out to be redundant. It is a ``secondary constraint'', which does not impose any further restriction on the fields. One can show explicitly that it follows from the point particle condition and its time derivative, but we can also use the results of the previous section. There we found, provided that the Einstein equations are satisfied, that the momentum of the particle is given by \eref{prt-hol-bnd}, and that the time component of the dreibein is given by a similar expression \eref{bnd-ee-t},
\beq
  \bar\mom = \bar\ggg^{-1} \mom \, \bar\ggg, \qquad
  \bar\eee_t = \del_t \tau \, \bar\ggg^{-1} \mom \, \bar\ggg. 
\eeq
Obviously, this implies that the time component of the dreibein on the boundary is proportional to the representation of the momentum vector in the local frames. Or, that the velocity of the particle is proportional to its momentum. The only extra information contained in \eref{bnd-sec} is that the physical time elapsing on the world line of the particle is determined by the value of the multiplier $\mult$ of the mass shell constraint.

\subsec{The fall off condition}

Let us now return to the ``boundary at infinity''. The problem is somewhat more complicated there, because, as it stands, the Lagrangian will in general not converge at $r\to\infty$. Let us write it explicitly as a limit, ignoring the particle term for a moment,
\beq[lag-lim]
 \fl \lag = \lim_{\rlim\to\infty}  \lagint , \qquad
  \lagint = \inti{r\le\rlim} \d^2x \, \eps^{ij} \, 
        \case12 \Trr{ \dot\omg_i \eee_j + 
               \omg_t \covD_i \eee_j +  \case12 \eee_t  \flds_{ij} }.
\eeq
We can now look for a minimal condition for this to converge, demanding some additional properties of this fall off condition. Thereby we shall, somewhat implicitly, use that we are dealing with a topological field theory.
As there are no local physical degrees of freedom carried by the gravitational field, it will be possible to localize the total energy, or the total action of a physical field configuration, that is, a solution to the field equations, within a compact region of the space manifold $\spce$. As a consequence, the total action of any field configuration satisfying the Einstein equations is always finite. We do not need to impose any \emph{further} restrictions. All we need to do, in order to get a well defined action principle, is to restrict the \emph{off shell} field configurations to those that still have a finite action.

If we also demand that the condition has the usual features, namely, that it only restricts the behaviour of the fields at infinity, and that it does not involve time derivatives, then it is, quite surprisingly, already fixed. It must be a ``subset'' (in the logical sense) of the field equations, as otherwise it would restrict their solutions. Actually, it must be a subset of the constraints, because the other field equations contain time derivatives. But then it cannot impose any restriction on the multipliers $\eee_t$ and $\omg_t$, because these are not involved in the constraints.

Now, consider the last two terms in the integrand above, which are proportional to the multipliers. If we want them to converge for \emph{any} choice of $\eee_t$ and $\omg_t$, then the constraints themselves must have a compact support. Hence, the fall off condition must read
\beq[fall-off-1]
  \exists \rlim \txt{such that, for $r\ge\rlim$,}
   \eps^{ij} \flds_{ij} = 0, \quad \eps^{ij} \covD_i \eee_j = 0 .
\eeq
We can write this in a slightly different way. Similar to the full Einstein equations, the general solution to the constraints can be given in terms of an embedding $(\ggg,\fff)$. It is formally the same as \eref{gen-sol}, except that now we have to replace the spacetime indices $\mu$ by space indices $i$. And, as in the previous section, we have to cut the region $r\ge\rlim$, which has the shape of a ring, along some radial line, say, again at $\p=\pm\pi$. As a result, the fall off condition becomes
\beq[fall-off-2]
   \exists \rlim \txt{such that, for $r\ge\rlim$,}
   \omg_i = \ggg^{-1}\del_i \ggg , \quad
   \eee_i = \ggg^{-1}\del_i \fff \, \ggg . 
\eeq
Along the cut at $\p=\pm\pi$, the fields $\ggg$ and $\fff$ are again denoted by $\ggg_\pm$ and $\fff_\pm$, and they have to satisfy a boundary condition like \eref{prt-emb-cut},
\beq[inf-emb-cut]
  \ggg_+ = \hol^{-1}  \ggg_- , \qquad
  \fff_+ = \hol^{-1}  (\fff_- - \ang) \, \hol.
\eeq
Note, however, that the situation is somewhat different from the one in the previous section. It is now only the space manifold $\spce$, and in fact only a ``neighbourhood of infinity'', which is embedded into Minkowski space. The fall off condition only requires that such an embedding exists at each moment of time, but it does not say anything about how the embedding at one time is related to that at another time. In particular, the ``constants'' $\hol\in\grpSL(2)$ and $\ang\in\algsl(2)$ can be arbitrary functions of time.

There is also a small but important difference between the two formulations of the fall off condition, \eref{fall-off-1} and \eref{fall-off-2}. Both can be thought of as a kind of ``asymptotical flatness'' condition. They state that it is possible to embed a neighbourhood of spatial infinity into flat Minkowski space. But in the second form, such an embedding is given \emph{explicitly}. Now, remember that the embedding $(\ggg,\fff)$ was determined by the dreibein and the spin connection up to a ridig Poincar\'e transformation only. Therefore, the fields $\ggg$ and $\fff$ contain more information than the dreibein and the spin connection.

If we use the second form \eref{fall-off-2} as the fall off condition, and regard $\ggg$ and $\fff$ as additional variables, then we add six new degrees of freedom to the system, corresponding to the six dimensions of the Poincar\'e group. At this stage, these are pure gauge degrees of freedom, because the action does not depend on them. However, we shall see in a moment that, in order to get a finite action, we are forced to add boundary terms to the action. They will explicitly depend on $\ggg$ and $\fff$. As a consequence, the six new degrees of freedom associated with the embedding will become physical. The action will no longer be gauge invariant under Poincar\'e transformations of the embedding Minkowski space.

There is also a natural physical interpretation of these extra degrees of freedom, which is typical for all kinds of asymptotical flatness conditions. We can think of the background frame at infinity as the rest frame of some observer. What we are effectively doing if we consider the embedding of spatial infinity as an additional degree of freedom of our model, is to add an external observer to the universe. A Poincar\'e transformation acting on $\ggg$ and $\fff$ then corresponds to a combination of rotations, boosts, and translations of the observer's reference frame. So far, it is however not clear whether this interpretation makes sense. What we have to show is that the observer ``couples'' correctly to the rest of the universe. In particular, the reference frame should stay fixed under the time evolution, but so far we do not have any equations of motion for the embedding.

\subsec{The boundary action at infinity}

To render the kinetic term in the Lagrangian \eref{lag-lim} finite, we cannot impose any further restriction on the fields at infinity, because we would then restrict the solutions to the field equations. The only way out is to regularize the action by adding a compensating boundary term. It is not obvious that such a term exists, but we shall see in a moment that it does. Let us compute the derivative of $\lagint$ with respect to $\rlim$, assuming that $\rlim$ is large enough, such that the fall off condition holds for this $\rlim$. Of course, it then also holds for every larger $\rlim$, and we can neglect the constraint terms,
\beq
\fl \frac{\del\lagint}{\del\rlim} = 
   \deldel{\rlim}  \Big(\inti{r\le\rlim} \d^2x \, \eps^{ij} \, 
                       \case12 \Trr{ \dot\omg_i \eee_j } \Big) = 
   \inti{r=\rlim} \d\p \, \case12 
               \Trr{ \dot\omg_r \eee_\p - \dot\omg_\p \eee_r }.
\eeq
Inserting \eref{fall-off-2}, and using the general formula $\del(\ggg^{-1}\delta\ggg) = \ggg^{-1}\delta(\del\ggg\,\ggg^{-1})\,\ggg$, which holds for any two commuting derivations $\del$ and $\delta$, this becomes
\beq
 \frac{\del\lagint}{\del\rlim} = 
   \inti{r=\rlim} \d\p \, 
    \case12 \Trr{ \del_r (\dot\ggg\,\ggg^{-1}) \del_\p \fff 
               - \del_\p (\dot\ggg\,\ggg^{-1}) \del_r \fff }.
\eeq
Here and in the following the integral over $\p$ always runs from $-\pi$ to $\pi$. If we now integrate the derivative $\del_\p$ in the last term by parts, we get
\beq
 \fl \frac{\del\lagint}{\del\rlim} =  
  \inti{r=\rlim} \d\p \, 
    \case12 \del_r \Trr{ \dot\ggg\,\ggg^{-1}  \del_\p \fff  }
  + \case12 \Trr{ \dot\ggg_- \, \ggg_-\inv \del_r \fff_- 
        - \dot\ggg_+ \, \ggg_+\inv \del_r \fff_+} . 
\eeq
Using the boundary condition \eref{inf-emb-cut}, this can be simplified to 
\beq
   \frac{\del\lagint}{\del\rlim} =  \deldel{\rlim} \Big(
    \inti{r=\rlim} \d\p \, \case12 \Trr{ \dot\ggg\,\ggg^{-1}\del_\p \fff}
    +  \case12 \Trr{ \hol^{-1} \dot \hol \, \fff_+ }  
        \Big). 
\eeq
From this we infer that the action becomes finite if we add the following boundary term,
\beq[lag-bnd]
  \lagbnd =  
       -  \inti{r=\rlim} \d\p \, 
              \case12 \Trr{ \dot\ggg\,\ggg^{-1}  \del_\p \fff  } 
        - \case12 \Trr{ \hol^{-1} \dot \hol \, \fff_+ } .
\eeq
Note that this is a kind of ``Wess Zumino Witten'' term that typically appears as a boundary action of a topological field theory \cite{Salomonson}. Taking everything together, the total Lagrangian becomes 
\beq[lag-tot]
     \lag = \lim_{\rlim\to\infty} (\lagprt + \lagint + \lagbnd),
\eeq
and it is now manifestly finite, because the fall off condition implies that, from some minimal $\rlim$ on, the expression between the parenthesis no longer depends on $\rlim$. The convergence is therefore quite strong, and we can evaluate the Lagrangian by taking $\rlim$ large enough. We can however not drop the limit, because the minimal value for $\rlim$ depends on the field configuration.

\subsec{The phase space}

We now have a well defined Lagrangian for our model, but so far we haven't shown that it provides the correct equations of motion at spatial infinity. This is somewhat involved, mainly due to the relations between the gravitational fields and the embedding at spatial infinity. The fall off condition restricts the possible variations of the fields in a non-trivial way, and these restrictions have to be taken into account when we derive the equations of motion. The actual calculation is most convenient carried out at the phase space level. We can then also use some of the results later on for the phase space reduction.

From the first order structure of the Lagrangian, it is more or less obvious how the basic fields split into phase space variables and Lagrange multipliers. The phase space is spanned by the spatial components of the gravitational fields, $\eee_i$ and $\omg_i$, and the embedding fields at spatial infinity, $\ggg$ and $\fff$. Let us denote these fields collectively by $\config$, and call the set of all such field configurations, subject to the ADM condition and the fall off condition, the \emph{extended phase space} $\exps$. From this, we shall later derive the \emph{reduced phase space} $\phsp$ as the quotient space of $\exps$ modulo gauge transformations. The total Lagrangian can then be split into a kinetic term,
\beq[lag-kin]
\fl  \lagkin = 
  \inti{r\le\rlim} \d^2x \, \eps^{ij} \, \case12 \Trr{ \dot \omg_i \, \eee_j  }
 - \inti{r=\rlim} \d\p \, \case12 \Trr{ \dot \ggg \, \ggg^{-1} \del_\p \fff} 
   - \case12 \Trr{\hol^{-1}  \dot \hol \, \fff_+ } ,
\eeq
and the Hamiltonian
\beq[ham]
   \ham = \mult ( \case12 \Trr{\bar\hol} - \cos \mass ) 
    \aln- \inti{r=0} \d\p \,  \case12 \Trr{\bar\omg_t  \bar\eee_\p}  \nwl
    \aln- \inti{r\le\rlim} \d^2x \, \eps^{ij} \, 
     \case12 \Trr{\omg_t  \covD_i \eee_j + \case12 \eee_t  \flds_{ij} } . 
\eeq
Here, we did not write out the limit $\rlim\to\infty$ explicitly, but it has to be understood in all what follows. The kinetic Lagrangian is thus a function of the phase space variables $\config$ and their velocities $\dot\config$. As it is linear in the velocities, we can directly read off the symplectic potential on $\exps$. It is given by
\beq[ex-sym-pot]
\fl  \pot = 
     \inti{r\le\rlim} \d^2x \, \eps^{ij} \, \case12 \Trr{ \dd \omg_i \, \eee_j  }
       - \inti{r=\rlim} \d\p \, \case12 \Trr{ \dd \ggg \, \ggg^{-1} \del_\p \fff} 
       - \case12 \Trr{\hol^{-1}  \dd \hol \,  \fff_+ } ,
\eeq
where $\dd$ denotes the exterior derivative on $\exps$. Acting with another exterior derivative on this, we find the symplectic form
\beq[ex-sym]
  \sym = \dd \pot  \aln=
    -  \inti{r\le\rlim} \d^2x \, \eps^{ij} \, 
                        \case12 \Trr{ \dd\omg_i \, \dd\eee_j }
    \nwl \aln{}{} 
   + \inti{r=\rlim} \d \p \,  
         \case12 \Trr{ \dd \ggg \, \ggg^{-1} \del_\p \dd\fff
           - \dd \ggg \, \ggg^{-1}\dd \ggg \, \ggg^{-1} \del_\p \fff} 
    \nwl \aln{}{} 
    + \vphantom{\int} \case12 
    \Trr{\hol^{-1}  \dd \hol \, \hol^{-1}  \dd \hol \, \fff_+ 
               + \hol^{-1}  \dd \hol \, \dd \fff_+ }  .
\eeq
This looks somewhat complicated, but, as we shall see in a moment, it is a non-degenerate two-form on $\exps$. We could therefore, in principle, derive the Poisson bracket from it. But this turns out to be rather difficult. For example, the brackets of the fields $\ggg$ and $\fff$ will be non-local (see, e.g., the appendix in \cite{Matschull2}, where a very similar action is considered). It is much more convenient to stick to the symplectic structure as the basic object on the phase space, and give an indirect proof of the non-degeneracy. One way to do this is to show that the Hamiltonian evolution equations,
\beq[ham-evolve]
  \sym \contract \dot \config = \dd\ham,
\eeq
can be solved uniquely for the velocities of the fields $\config\in\exps$. The left hand side of this equation is obtained by contracting the symplectic form with the velocities,
\beq[sym-dot]
 \sym \contract \dot \config  \aln= 
     \inti{r\le\rlim} \d^2x \, \eps^{ij} \, 
      \case12 \Trr{ \dot \omg_i \, \dd \eee_j +  \dot \eee_i \,  \dd \omg_j  } 
   \nwl\aln{}{}
   +  \inti{r=\rlim} \d \p \, 
        \case12 \Trr{ \dot \omg_\p \,  \ggg^{-1} \dd \fff \, \ggg +
                    \dot \eee_\p \,  \ggg^{-1} \dd \ggg  } 
   \nwl\aln{}{}
   + \vphantom{\int}\case12 \Trr{ \dot \ggg_-  \ggg_-\inv  \dd\ang
              + (\dot \fff_+ + [ \fff_+ , \dot\ggg_+ \ggg_+\inv ] ) 
                           \,    \hol^{-1}  \dd\hol } .
\eeq
Here, we integrated several times by parts, made use of the boundary condition \eref{inf-emb-cut} along the cut at $\p=\pm\pi$, and we used that, for sufficiently large $r$, the derivatives of $\omg_i$ and $\eee_i$ are given in terms of those of $\ggg$ and $\fff$ by
\beq[del-emb]
  \dd\omg_i \aln= \dd(\ggg^{-1}\del_i\ggg) =
                \ggg^{-1} \del_i (\dd\ggg \, \ggg^{-1} ) \, \ggg, \nwl
  \dd\eee_i \aln= \dd(\ggg^{-1}\del_i\fff\,\ggg) = 
            \ggg^{-1} \del_i \dd \fff \, \ggg +
            [\ggg^{-1}\del_i\fff\,\ggg,\ggg^{-1}\dd\ggg].
\eeq
Similarly, the exterior derivative of the Hamiltonian is found to be 
\beq[dd-ham]
 \fl \dd\ham  \aln= 
    \inti{r=0} \d\p \, 
        \case12 \Trr{ ( \bar\eee_t - \mult \, \bar\mom )\, \dd \omg_\p } 
     \nwl\aln{}{} 
   + \inti{r\le\rlim} \d^2x \, \eps^{ij} \, 
     \case12 \Trr{ \covD_i \omg_t \, \dd \eee_j +
           ( \covD_i \eee_t + [\eee_i,\omg_t] ) \, \dd \omg_j  } 
     \nwl\aln{}{}
   +  \inti{r=\rlim} \d \p \, 
     \case12 \Trr{ \covD_\p \omg_t \, \ggg^{-1} \dd \fff \, \ggg +
           ( \covD_\p \eee_t + [\eee_s,\omg_t] ) \, \ggg^{-1} \dd \ggg  } 
     \nwl\aln{}{}
   + \vphantom{\int} \case12 \Trr{ \ggg_- \omg_t \, \ggg_-\inv  \dd\ang
              + (\ggg_+ \eee_t \, \ggg_+\inv 
                 + [ \fff_+ , \ggg_+\omg_t\,\ggg_+\inv ]  ) 
                        \,   \hol^{-1}  \dd\hol }   . 
\eeq
Now, we can read off the evolution equations, and the secondary constraints on the particle boundary. The latter follows from the first line in \eref{dd-ham}, which has no counterpart in \eref{sym-dot}. It is exactly the second equation in \eref{bnd-sec}. To recover the evolution equations for $\omg_i$ and $\eee_i$, we have to compare the integrals over the range $r\le\rlim$ in \eref{sym-dot} and \eref{dd-ham}. Obviously, they are still the same as \eref{evolve}, and they must be satisfied \emph{everywhere} on $\spce$, because the two integrals must be equal in the limit $\rlim\to\infty$.

To find the remaining equations of motion, let us assume that those already obtained are satisfied. We are then left with the last lines in \eref{sym-dot} and \eref{dd-ham}, containing terms proportional to the exterior derivatives of $\hol$ and $\ang$. They also contain the time derivatives of the fields $\ggg$ and $\fff$, so we expect them to provide an evolution equation for the embedding. Now, the time evolution of the embedding is already almost fixed by that of the dreibein and the spin connection. The only remaining freedom is to perform a rigid Poincar\'e transformation. Explicitly, we find that
\beq[emb-evolve-gen]
  \dot \ggg = \ggg \, \omg_t - \rlgen \, \ggg, \qquad
  \dot \fff = \ggg \, \eee_t \, \ggg^{-1} - [\rlgen,\fff] - \rtgen,
\eeq
where the parameters $\rlgen,\rtgen\in\algsl(2)$ are spatially constant, but may be chosen freely at each moment of time. This freedom corresponds to an infinitesimal generator of a Poincar\'e transformation \eref{emb-pnc}. Without the boundary terms in the action, $\rlgen$ and $\rtgen$ would provide six free parameters, and the embedding would be a gauge degree of freedom. But now we get six extra equations of motion as well, which fix the constants $\rlgen$ and $\rtgen$. Inserting \eref{emb-evolve-gen}, the difference of the last lines in \eref{sym-dot} and \eref{dd-ham} becomes a rather simply expression,
\beq
 \case12 \Trr{ \rlgen \, \dd \ang +
                \rtgen \, \hol^{-1} \dd\hol}   .
\eeq
This vanishes if and only if $\rlgen=0$ and $\rtgen=0$, because $\hol$ and $\ang$ are \emph{independent functions} on the phase space $\exps$. They are implicitly defined by the relations \eref{inf-emb-cut}, and it is easy to check that their values are not restricted. The time evolution of the embedding of spatial infinity is therefore fixed, and we have
\beq[emb-evolve]
  \ggg^{-1} \dot \ggg = \omg_t , \qquad
  \ggg^{-1} \dot \fff \, \ggg = \eee_t. 
\eeq
Together with the fall off condition \eref{fall-off-2}, we recover the three dimensional equations \eref{gen-sol}. As a result, the frame at infinity behaves such that it provides an embedding of \emph{spacetime} into Minkowski space, although it has been defined as an embedding of \emph{space} only. We can also say that the evolution equations are such that the observer at infinity is correctly ``coupled'' to the universe. His reference frame is not shifted or rotated under the time evolution.

Moreover, we were able to solve the Hamiltonian equations of motion uniquely for the velocities of the phase space variables $\dot\config$, for any initial field configuration $\config$. This implies that the symplectic form \eref{ex-sym} is non-degenerate, as otherwise the Hamiltonian equations \eref{ham-evolve} could not be solved for $\dot\config$. Without explicitly deriving the inverse of $\sym$, that is, the Poisson bracket, we know that we identified the phase space structure of the system correctly.

\subsec{Gauge symmetries}

Finally, let us very briefly discuss the gauge symmetries of our model, which we want to divide out in the next section. The basic gauge symmetries of Einstein gravity in the dreibein formulation are the local Poincar\'e transformations,
\beq[loc-pnc]
  \delta \omg_\mu = \covD_\mu \llgen , \qquad
  \delta \eee_\mu = \covD_\mu \ltgen + [\eee_\mu,\llgen].
\eeq
Here, $\llgen\in\algsl(2)$ is the generator of a local Lorentz rotation, and $\ltgen\in\algsl(2)$ that of a local translation. For the fall off condition to be preserved, the embedding fields have to transform according to
\beq[loc-pnc-emb]
  \delta\ggg = \ggg\, \llgen , \qquad
  \delta\fff = \ggg\, \ltgen \, \ggg^{-1}.
\eeq
The generator of a spacetime diffeomorphism is obtained, at least on shell, by fixing a vector field $\dfgen^\mu$ and choosing the parameters to be $\llgen=\dfgen^\mu\omg_\mu$ and $\ltgen=\dfgen^\mu\eee_\mu$. Then we have
\beq[diff]
  \delta\eee_\mu \aln= 
   \dfgen^\nu \, \del_\nu \eee_\mu + \del_\mu \dfgen^\nu \, \eee_\nu 
         + 2\, \dfgen^\nu \, \covD_{[\mu}\eee_{\nu]},   
   \nwl 
  \delta\omg_\mu \aln= 
   \dfgen^\nu \, \del_\nu \omg_\mu + \del_\mu \dfgen^\nu \, \omg_\nu 
          + \dfgen^\nu \, \flds_{\mu\nu}.
\eeq
Up to terms proportional to the field equations, these are the Lie derivatives of the one-forms. Similarly, one finds that the same holds for the scalars $\ggg$ and $\fff$. It is also straightforward to check that all these transformations are symmetries of the action, provided that, on the particle boundary, the parameter $\ltgen$ obeys
\beq[bnd-gen-con]
  \bar\ltgen = \msgen \, \bar\mom,
\eeq
for some real number $\msgen$, and that one includes a transformation of the multiplier of the mass shell constraint,
\beq[loc-pnc-prt] 
   \delta \mult = \dot\msgen.
\eeq
Of course, this restriction arises in the same way as the secondary constraint for the time component of the dreibein on the particle boundary \eref{bnd-sec}. It is a result of the restricted diffeomorphisms invariance on the boundary. If we write \eref{bnd-gen-con} as a restriction on the generating vector field $\dfgen^\mu$ of a diffeomorphism, then the condition becomes $\bar\dfgen^r=0$ and $\del_\p\bar\dfgen^t=0$. The first condition is quite reasonable. It states that the vector field is tangent to the boundary, which means that the diffeomorphism maps the boundary onto itself. The second condition implies that the diffeomorphism does not mix the circles of constant $t$ on the boundary.

In other words, we do not have the full diffeomorphism group on the boundary as a gauge group. This is because the circles of constant $t$ play a distinguished role. They represent a \emph{single physical point} on the world line of the particle. It is the point particle equation $\bar\eee_\p=0$, which explicitly refers to these circles, and brakes the diffeomorphism invariance. Off the boundary however the diffeomorphism group is unbroken, and there is a one-to-one relation between local translations and diffeomorphisms, because the dreibein, which relates the generating vector fields $\dfgen^\mu$ and $\ltgen$, is invertible. We should also note that the fall off condition is diffeomorphism invariant, although it refers to the special coordinate $r$. It is equivalent to the manifestly invariant statement that there exists an embedding outside some compact subset of $\spce$.

At the phase space level, the local Poincar\'e transformations behave as it is expected. The transformation of the phase space variables $\eee_i$, $\omg_i$, $\ggg$ and $\fff$ are determined by the parameters $\llgen$ and $\ltgen$ alone, whereas those of the multipliers $\eee_t$, $\omg_t$ and $\mult$ depend on the time derivatives of the parameters as well. We can also show that the gauge transformations $\delta\config$ on the phase space $\exps$ are generated by the constraints. We have $\delta\config=\{\con,\config\}$, or equivalently $\sym\contract\delta\config = \dd\con$, if we set
\beq
 \con = 
        \msgen \, ( \case12 \Trr{\bar\hol} - \cos \mass ) 
   \aln- \inti{r=0} \d\p \, \case12 \Trr{\bar\llgen \, \bar\eee_\p} \nwl
   \aln- \inti{r\le\rlim} \d^2x \, \eps^{ij} \, 
     \case12 \Trr{ \llgen \, \covD_i \eee_j + \case12 \ltgen \, \flds_{ij} } .
\eeq
This is, like the Hamiltonian, a linear combination of the constraints. But it is not the most general linear combination. The parameter of local translations on the boundary is restricted by \eref{bnd-gen-con}. It is however the most general linear combination that is \emph{first class} and thus a generator of a gauge symmetry.

That some of the constraints are second class can also be inferred from the fact that there are secondary constraints which restrict the associated multipliers, and therefore the gauge symmetries. An explicit derivation of the Poisson bracket on the extended phase space $\exps$ would therefore not be of much use. For the derivation of the reduced phase space, we had to go over to the Dirac bracket anyway, and this would make things even more complicated. If we however stick to the symplectic formalism, the corresponding structure on the reduced phase space can be found quite easily, by pulling back the symplectic form $\sym$ onto the constraint surface, and then dividing out the gauge degrees of freedom, which happens almost automatically.

Finally, let us briefly come back to the gauge transformations \eref{loc-lor} defined in the beginning as the most general transformations of the local frames that leave the metric and the time orientation of spacetime invariant,
\beq[loc-lor-fin]
  \eee_\mu \mapsto \llpar^{-1} \eee_\mu \, \llpar , \qquad
  \omg_\mu \mapsto \llpar^{-1} (\del_\mu + \omg_\mu) \, \llpar .
\eeq
This is the finite version of a local Lorentz transformation. There is however a crucial difference between the infinitesimal transformations above and these finite ones. Some of the transformation given by \eref{loc-lor-fin} are \emph{large} gauge transformations, which means that they are not generated by \eref{loc-pnc}.

As an example, consider the maps $\llpar=\expt{-z\p\gam_0}$, with $z\in\Zset$, as parameters of local Lorentz transformations. They are well defined, because $\expt{2\pi\gam_0}=\one$, and one can easily show that every map $\llpar:\spce\to\grpSL(2)$ can be smoothly deformed into one of these, but it is not possible to deform them into each other. So, the set of all maps $\llpar$, and therefore the gauge group, splits into a series of disconnected components, labeled by the ``winding number'' $z$. Only the transformations with $z=0$ are generated by the infinitesimal transformations \eref{loc-pnc}. 

To see what a large gauge transformations does, let us choose $\llpar=\expt{-z\p\gam_0}$ and act with this on the fields in the neighbourhood of the particle, as given by \eref{rst-ee} and \eref{rst-om}. The result is that the dreibein is replaced by 
\beq[wind-ee-1]
  \eee_t \aln = \gam_0 ,\qquad
  \eee_r = \cos((2z+1)\p) \, \gam_1 + \sin((2z+1)\p) \, \gam_2 ,
  \nwl 
  \eee_\p \aln = \Big( 1-\frac\pi\mass \Big) \, r \,
          \big( \cos((2z+1)\p) \, \gam_2 - \sin((2z+1)\p) \, \gam_1 \big),
\eeq
and the spin connection becomes
\beq[wind-om-1]
  \omg_t = 0 ,\qquad
  \omg_r = 0 ,\qquad
  \omg_\p = - \Big( \frac{\mass}{2\pi} + z \Big) \, \gam_0 .
\eeq
So, the new dreibein ``winds'' differently around the boundary at $r=0$, and the value of the angular component of the spin connection has changed as well. But, for example, the holonomy of the particle is still the same,
\beq[wind-hol-1]
  \bar\hol = \expo{ (\mass+2\pi z) \gam_0 } = \expo{\mass\gam_0},
\eeq
and the metric is of course also unchanged. There is no way to distinguish the field configurations for different winding numbers $z$ physically, if we are only allowed to refer to the metric, and to use vector or spinor test particles to measure holonomies. This is the reason why we have to consider the large Lorentz transformations as gauge transformations. What seems a bit strange, however, is that the winding number of the dreibein in \eref{wind-ee-1} is always odd. The following dreibein represents the same metric,
\beq[wind-ee-2]
  \eee_t \aln = \gam_0 ,\qquad
  \eee_r = \cos(2z\p) \, \gam_1 + \sin(2z\p) \, \gam_2 ,
  \nwl 
  \eee_\p \aln = \Big( 1-\frac\pi\mass \Big) \, r \,
          \big( \cos(2z\p) \, \gam_2 - \sin(2z\p) \, \gam_1 \big),
\eeq
but it is not possible to transform it into \eref{wind-ee-1}. What is the difference between these field configurations?  To find the answer, we have to evaluate the spin connection,
\beq[wind-om-2]
  \omg_t = 0 ,\qquad
  \omg_r = 0 ,\qquad
  \omg_\p = - \Big( \frac{\mass-\pi}{2\pi} + z \Big) \, \gam_0 ,
\eeq
and compute the holonomy
\beq[wind-hol-2]
  \bar\hol = \expo{ (\mass - \pi + 2\pi z ) \gam_0 } 
            = \expo{ -(\pi-\mass) \gam_0 } = - \expo{ \mass\gam_0 } .
\eeq
This is not the same as \eref{wind-hol-1}. The two particles can be distinguished physically, by transporting a spinor around them. The resulting spinors will differ by an overall sign. The second particle is, using the terminology of the previous section, an \emph{antiparticle} with mass $\pi-\mass$. We have seen that this has the same deficit angle as a particle with mass $\mass$, so that the metric in the neighbourhood is the same.

From this we conclude, and this will be rather important later on, that the whole physical information about the particle is encoded in the \emph{spinor} representation of the holonomy $\bar\hol\in\grpSL(2)$. It is not sufficient to consider the vector representation, because then we cannot distinguish between particles and antiparticles. But on the other hand, it is also not appropriate to consider its representation in any ``higher'' covering of $\grpSL(2)$ as a physical quantity. This would no longer be invariant under general local Lorentz transformations, because the last equality in \eref{wind-hol-1} no longer holds. This is why $\grpSL(2)$ will become to the ``momentum space'' of the particle in the next section, and not a covering thereof.

\section{Phase space reduction} 
\label{reduce}

When quantizing a gauge theory, there are two roads between which one has to choose. Either one can quantize the full theory, including gauge degrees of freedom, and use operator constraint equations on the Hilbert space, or one can divide out the gauge degrees of freedom at the classical level, and then quantize. In the case of a topological field theory, the number of physical degrees of freedom that remain after such a phase space reduction is finite, and therefore it is more convenient to first reduce and then quantize. This is the road we will follow in this article. We shall however, in a first step, not divide out all gauge symmetries. Instead, we only consider those that are generated by the ``kinematical'' constraints,
\beq[kin-con]
  \eps^{ij} \flds_{ij} = 0 , \qquad
  \eps^{ij} \covD_i\eee_j = 0 , \qquad
  \bar\eee_\p = 0.
\eeq
If we do so, then there is only one constraint left, the mass shell constraint, and this will become a ``dynamical'' constraint, in the sense that it generates the time evolution of the particle seen by the observer at infinity. The resulting \emph{reduced phase space} $\phsp$ will, strictly speaking, only be a partially reduced phase space, but it will already be finite dimensional.

With the preparation made in the previous section, the actual phase space reduction is quite simple. The first thing we need is an appropriate set of coordinates on the ``constraint surface'', which is defined by \eref{kin-con} as a subset of the extended phase space $\exps$. This is not very complicated. All we need to do is to take the embedding of spatial infinity $(\ggg,\fff)$, and extend it to the whole space manifold, such that, everywhere on $\spce$, we have
\beq[con-sol]
  \omg_i = \ggg^{-1} \del_i \ggg, \qquad
  \eee_i = \ggg^{-1} \del_i \fff \, \ggg.
\eeq
As we already found in \sref{particle}, the point particle constraint implies that $\fff$ is constant on the particle boundary. There is thus a one-to-one relation between field configurations satisfying the kinematical constraints and embeddings of $\spce$, cut along the radial line at $\p=\pm\pi$, with the condition that $\bar\fff=\xpos$ is independent of $\p$.

Using the variables $\ggg$ and $\fff$ as coordinates, we can compute the pullback of the symplectic form $\sym$ on the constraint surface. It is actually simpler to first compute the pullback of the symplectic potential $\pot$, and then use that $\sym=\dd\pot$. All we need to do is to insert \eref{con-sol} into the expression \eref{ex-sym-pot} for $\pot$. Thereby, we can use the following trick. We know that the full expression is independent of $\rlim$, provided that $\rlim$ is large enough, which means that the fall off condition \eref{fall-off-2} must hold. But now, this is fulfilled for any $\rlim$. In particular, we can take $\rlim=0$. Doing so, the first integral vanishes because the integration range shrinks to a set of measure zero. The second integral also vanishes, because $\bar\fff=\xpos$ is constant and therefore $\del_\p\fff=0$ at $r=0$. Only the last term remains, and there we can replace $\fff_+$ with $\xpos$, because this is now also evaluated at $r=0$. All together, this gives
\beq[pot-red]
   \pot = - \case12 \Trr{\hol^{-1}\dd\hol\,\xpos}.
\eeq
The Hamiltonian also takes a rather simple form. As all constraints except the mass shell constraint have been solved, this is the only term that remains,
\beq[ham-red]
  \ham =  \mult\, ( \case12 \Trr{\hol} - \cos\mass ).
\eeq
Here, we also used the relation $\bar\hol=\bar\ggg^{-1}\hol\,\bar\ggg$, to replace $\bar\hol$ with $\hol$. Thus, the only variables that remain are $\xpos\in\algsl(2)$ and $\hol\in\grpSL(2)$. These are the only ``observables'', that is, the only quantities that are invariant under the gauge transformations generated by the kinematical constraints. We can also reexpress everything in terms of a reduced Lagrangian, which only depends on these variables,
\beq[lag-red]
    \lag = - \case12 \Trr{\hol^{-1}\dot\hol\,\xpos} 
              - \mult\, ( \case12 \Trr{\hol} - \cos\mass ).
\eeq
To check our calculations, we can rederive the equations of motion from this. Of course, the multiplier $\mult$ still provides the correct mass shell condition. Under a variation of $\xpos$, we get
\beq
  \delta\lag = \case12 \Trr{ \hol^{-1} \dot \hol \, \delta\xpos } 
  \follows \dot\hol=0,
\eeq
and varying $\hol$ gives 
\beq
  \delta\lag = \case12 \Trr{ \delta(\hol^{-1}\dot\hol) \, \xpos  
                  + \mult \, \delta \hol }
             = \case12 \Trr{ \hol^{-1} \delta \hol \, 
                      ( \mult \, \mom - \dot\xpos )}.
\eeq
Here, we used $\delta(\hol^{-1}\dot\hol) = \hol^{-1}\del_t(\delta\hol\,\hol^{-1})\,\hol$, and $\dot\hol=0$. As a result, we find the evolution equations
\beq[red-evolve]
  \dot\hol = 0 , \qquad
  \dot\xpos = \mult \, \mom,
\eeq
which are in agreement with what we found in \sref{particle}. In particular, the momentum $\hol$ is constant, and the world line of the particle in the embedding space is given by \eref{prt-emb-bnd}, with the evolution of the eigentime $\tau$ given by the multiplier $\mult$.

The physical interpretation of the observables $\xpos$ and $\hol$ is now also rather obvious. We found that the value of $\fff$ on the particle boundary was the position of the particle in the embedding Minkowski space. This space is now identical with the reference frame of the observer at infinity, and therefore $\xpos$ is the \emph{position} of the particle, seen by this observer. Similarly, $\hol$, or its projection into the algebra $\mom$, is the \emph{momentum} of the particle, measured by the same observer. Together they span the reduced phase space $\phsp=\grpSL(2)\times\algsl(2)$.

We can now forget about the rather long derivation of reduced phase space $\phsp$ from Einstein gravity, and summarize the result as follows. The model effectively describes a point particle from the point of view of an external observer. It has many features in common with a relativistic point particle. The phase space $\phsp$ is six dimensional, with the basic variables being the position and the momentum of the particle, and the physical subspace of $\phsp$ is defined by a mass shell constraint. The dynamics of the particle will be described in the same way, namely as a gauge symmetry generated by the mass shell constraint.

There is however one crucial difference. The phase space $\phsp$ is not a six dimensional vector space, but the product of a group manifold with its Lie algebra, which is the same as the cotangent bundle of the group manifold, $\phsp=\tang_*\grpSL(2)$. Such phase spaces are well known. They typically appear as the phase spaces of, say, particles moving on a curved manifold, or field theories with curved target spaces. But here the situation is somewhat different. It is not the \emph{position} variable that lives on the curved manifold, but the \emph{momentum}, and this will lead to some very uncommon features of this model, especially in the quantum theory.

\subsec{Poisson brackets} 

Many of these features can already be seen at the classical level, and most of them follow immediately from the Poisson structure on $\phsp$. Exploiting the cotangent space structure, it is more or less straightforward to derive the Poisson bracket. Let us expand the position variable $\xpos$ in terms of the gamma matrices and rewrite the symplectic potential as
\beq
   \pot =  - \case12 \Trr{\hol^{-1} \d\hol\,\gam_a}\, \xposv^a.
\eeq
The one-form $\case12 \Trr{\hol^{-1} \d\hol\,\gam_a}$ is the dual of the left invariant vector field $\Lvec_a$ on the group manifold, which is defined by
\beq
     \Lvec_a \, \hol = \case12 \, \hol\gam_a.
\eeq
It is therefore reasonable to assume that the action of $\xposv^a$ under the Poisson bracket is associated with this vector field. With this ansatz, and using the Jacobi identity, it is not difficult to derive the basic Poisson brackets, which turn out to be
\beq[pois]
  \pois{\hol}{\xposv^a} = \hol\gam^a,\qquad
  \pois{\xposv^a}{\xposv^b} = 2\,\eps^{ab}\_c \, \xposv^c.
\eeq
An alternative way to derive this bracket, without using the cotangent space structure, is to consider the components $\momv_a$ and $\xposv^a$ as independent variables, and insert the expansion $\hol = \moms\,\one+\momv^a\,\gam_a$, with $\moms=\sqrt{\momv^a\momv_a+1}$, into the expression for the symplectic potential. It then becomes
\beq
  \pot = - \qposv^a \dd \momv_a , \qquad
  \qposv^a = ( \eta^{ab} \moms +  \eps^{abc} \momv_c - 
               \moms^{-1} \momv^a\momv^b ) \, \xposv_b.
\eeq
Hence, it has the canonical form, and the brackets can then be derived from the condition
\beq
  \pois{ \momv_a }{ \momv_b } = 0 ,\qquad
  \pois{ \momv_a }{ \qposv^b } = \delta_a\^b , \qquad
  \pois{ \qposv^a }{ \qposv^b } = 0 .
\eeq
The result is
\beq[pois-x-q]
  \pois{\momv_a}{\xposv^b} = 
    \delta_a\^b \, \moms + \eps_a\^{bc} \, \momv_c, \qquad
  \pois{\moms}{\xposv^b} = \momv^b.
\eeq
This also follows from \eref{pois}, by taking taking the trace or projecting it onto the gamma matrices, and vice versa it implies that $\xposv^a$ acts on $\hol$ by right multiplication with $\gam^a$. As a ``proof'', we rederive the time evolution equations from the Hamiltonian \eref{ham-red},
\beq[ham-pois]
  \dot \hol = \pois\ham\hol = 0 ,\qquad
  \dot \xposv^a = \pois\ham{\xposv^a} 
            = \case12 \mult  \, \Trr{\hol\gam^a} 
            =  \mult \, \momv^a.
\eeq
As $\mult$ is a free parameter, this is actually a gauge transformation, generated by the mass shell constraint. Once again, this is the same situation as for the relativistic point particle. It is also the reason why we did not divide out all gauge symmetries from the very beginning. We want to keep the remaining gauge freedom when we quantize the model, because we can then easily compare the result to the quantized relativistic point particle.

Before coming to this, we shall however stick to the classical description for a moment and see what the basic \emph{physical} properties of the reduced phase space $\phsp$ are. The most apparent feature of the Poisson algebra is of course that the components $\xposv^a$ of the position of the particle have non-vanishing brackets. The fact that the corresponding quantum operators do not commute will have profound consequences for the quantum theory. For example, they add another uncertainty relation to the quantum mechanics, which implies that the particle cannot be localized.

We should emphasize that this ``non-commutativity of spacetime'' is not the result of any choice. The interpretation of the ``observable'' $\xpos$ as the position of the particle with respect to the observer at infinity is forced upon us, as a consequence of the definitions made at the \emph{field theoretical} level in the beginning. At this point, the description of the particle was completely independent from that of the observer. The only connection between them was the gravitational field, and its dynamics is described by the Einstein Hilbert action. So, the fact that the position coordinates do not commute turns out to be an effect of Einstein gravity.

\subsec{Symmetries}

The last issues we want to investigate at the classical level are the symmetries and their Noether charges. A transformation is a symmetry if the Lagrangian is invariant up to a total time derivative. If this is the case, then the symmetry is realized on phase space as a canonical transformation, and we should find the associated Noether charge that generates the symmetry via the Poisson bracket. As a symmetry also leaves the Hamiltonian invariant, which in this case means that its generator commutes with the mass shell constraint, it represents an observable of the system. The algebra of the Noether charges is therefore simultaneously the observable algebra, which will play a central role in the quantum theory.

Let us first consider Lorentz transformations. It is easily checked that the reduced Lagrangian \eref{lag-red} is invariant under
\beq
 \hol \mapsto \rlpar^{-1} \hol \, \rlpar, \qquad 
 \xpos \mapsto \rlpar^{-1} \xpos \, \rlpar,
\eeq
where $\rlpar\in\grpSL(2)$ is a constant. At the extended phase space level, this corresponds to a ridig Lorentz transformation of the embedding Minkowski space at spatial infinity. Infinitesimally, the transformation can be written as
\beq[lor-gen]
 \delta \hol= \case12\llgenv^a \, [ \hol , \gam_a ] , \qquad 
 \delta \xpos= \case12\llgenv^a \, [ \xpos , \gam_a ].
\eeq
To find the Noether charge corresponding to this symmetry, we need a function $\angv_a$ satisfying
\beq[J-pois]
 \pois{\angv_a}{\hol} = \case12 [\hol,\gam_a],\qquad
 \pois{\angv_a}{\xpos} = \case12 [\xpos,\gam_a].
\eeq
The action of $\angv_a$ on $\hol$ is the sum of the left and right invariant vector fields, defined by
\beq[LR-fields]
  \Lvec_a \, \hol  = \case12 \, \hol \, \gam_a ,\qquad
  \Rvec_a \, \hol = - \case12 \, \gam_a \hol.
\eeq
They form two commuting copies of an $\algso(1,2)$ Lie algebra,
\beq[LR-comm]
  [ \Lvec_a , \Lvec_b ] = \eps_{ab}\^c \, \Lvec_c, \quad
  [ \Rvec_a , \Rvec_b ] = \eps_{ab}\^c \, \Rvec_c, \quad
  [ \Lvec_a , \Rvec_b ] = 0.
\eeq
Moreover, the right invariant vector fields can be written as $\Rvec_a=-\holv_a\^b\Lvec_b$, where $\holv_a\^b$ is the vector representation of $\hol$ (see \eref{SL-SO}). Using this, it is not difficult to identify the angular momentum as
\beq[Ja]
  \angv_a = \case12 ( \holv_a\^b \xposv_b - \xposv_a ).
\eeq
As expected, its components form an $\algso(1,2)$ algebra under the Poisson bracket,
\beq
 \pois{\angv_a}{\angv_b} = -\eps_{ab}\^{c}\, \angv_c.
\eeq
If we write it as a matrix, we find that the angular momentum is, up to a factor, the same as the parameter $\ang$ appearing in the boundary condition \eref{prt-emb-cut}, which is related to the position and momentum of the particle by \eref{prt-ang},
\beq
  \angv^a \, \gam_a = \case12 ( \holv^a\_b \, \xposv^b - \xposv^a ) \, \gam_a 
              = \case12 ( \hol \, \xpos \, \hol^{-1} - \xpos )
              = - \case12 \, \ang.
\eeq
We can also express $\angv_a$ as a function of the components of the position and the momentum vector. It then becomes
\beq[JxQ]
  \angv_a =  \moms\, \eps_{abc} \, \xposv^b \momv^c 
     + 2 \, \xposv_{[a} \momv_{b]} \, \momv^b.
\eeq
This expression is different from what one would expect naively, namely $\eps_{abc}\,\xposv^b\momv^c$. Only in the ``low momentum'' limit, where $\momv_a$ is small compared to $\moms\approx1$, it reduces to the standard expression.

If we look at the action of $\angv_a$ and $\xposv_a$ on $\hol$, we find that there is a remarkable similarity. If $F(\hol)$ is some function on $\grpSL(2)$, then we have
\beq
  \pois{\xposv_a}{F} = - 2\,\Lvec_a \, F, \qquad
  \pois{\angv_a}{F} = (\Lvec_a+\Rvec_a) \, F. 
\eeq
The left and right invariant vector fields $\Lvec_a$ and $\Rvec_a$ span an $\algso(1,2)\times\algso(1,2)\simeq\algso(2,2)$ Lie algebra. This is the symmetry algebra of anti-de-Sitter space. Indeed, at least locally, our momentum space looks like anti-de-Sitter space. If we use the coordinates $\momv_A=(\momv_a,\moms)$, then they obey
\beq
  \momv_A \momv_B \eta^{AB} = -1, \qquad  \eta^{AB}=\diag(-1,1,1,-1).
\eeq
Hence, it is the unit hyperboloid in $\Rset^{2,2}$, and its isometries are generated by the vector fields
\beq
  \ads_{AB} = \case12 \Big( \momv_A \deldel{\momv^B} 
                        - \momv_B \deldel{\momv^A} \Big).
\eeq
If we split them into 
\beq
  \ads_{ab} =  \case12 \Big( \momv_a \deldel{\momv^b} 
                         - \momv_b \deldel{\momv^a} \Big) ,
            \qquad
  \ads_{a} =  \case12 \Big( \moms \deldel{\momv^a} 
                        + \momv_a \deldel{\moms} \Big),
\eeq
then it is not difficult to verify that
\beq
  \pois{\xposv_a}{F} = -2 \,\ads_{a} \, F - \eps_a\^{bc}\, \ads_{bc}\, F , 
    \qquad
  \pois{\angv_a}{F} =  \eps_a\^{bc} \, \ads_{bc}\, F,
\eeq
where $F(\hol)$ is considered as a function of the components $\momv_A$. Hence, $\xposv_a$ and $\angv_a$ generate the isometries of momentum space. This is again analogous to the relativistic point particle. There, the momentum space is Minkowski space, its isometry group is $\grpISO(1,2)$, and the components of the position and angular momentum vectors are the generators of translations and Lorentz transformations. Here, the situation is exactly the same, except that the isometry group of momentum space is $\grpSO(2,2)$. This is the reason why the algebra of the position and angular momentum components is different.

The idea of momentum space being anti-de-Sitter is actually not new. It was proposed in 1946(!)  by Snyder in the context of relativistic point particles in 3+1~dimensions \cite{Snyder}. He realized that this implies that the components of the position vector of a particle can no longer commute with each other. Quite remarkably, it also introduces a shortest distance into the quantum theory, without breaking the Lorentz symmetry, as it typically happens when a lattice structure is introduced on spacetime. In other words, it is possible to introduce a cutoff in spacetime in a covariant way. At the quantum level, we shall find that our model has all these features as well.

So far, we considered the isometries of the \emph{momentum} space. But what about those of the \emph{position} space, or the \emph{spacetime} in which the particle lives. As this is flat Minkowski space, we should expect that there is an $\grpISO(1,2)$ symmetry group, and using the analogy to the relativistic point particle, the symmetries should be generated by the components of the momentum $\momv_a$ and the angular momentum $\angv_a$. Indeed, these quantities form a Poincar\'e algebra,
\beq
      \pois{\momv_a}{\momv_b} = 0, \quad
      \pois{\angv_a}{\momv_b} = - \eps_{ab}\^c \momv_c, \quad
      \pois{\angv_a}{\angv_b} = - \eps_{ab}\^c \angv_c. 
\eeq
We know already that $\angv_a$ is the Noether charge associated with Lorentz transformations \eref{J-pois}. So, we should expect that $\momv_a$ is the generator of a translation of the particle. Let us contract the momentum with an arbitrary vector $\dtgenv^a$, and compute the bracket with the position coordinate. What we get is
\beq
   \pois{\dtgenv^b \momv_b}{\xposv^a} 
     = \moms \, \dtgenv^a - \eps^{abc}\, \dtgenv_b\momv_c .
\eeq
As the $\momv$'s themselves have vanishing brackets with each other, we can easily exponentiate this transformation, and write down the action of the symmetry on the basic phase space variables, parametrized by a vector $\dtgen=\dtgenv^a\gam_a$,
\beq[translation]
   \hol \mapsto \hol, \qquad
   \xpos \mapsto \xpos + \moms \, \dtgen + [\dtgen,\mom] .
\eeq
This does not look like an ordinary translation $\xpos\mapsto\xpos+\dtgen$. Maybe the reason for this is that the components of the momentum $\mom$ are just not the correct generators. To see that this is not the case, consider the transformation of the Lagrangian \eref{lag-red} under an ordinary translation. It is given by
\beq[lag-shift-1]
   \lag \mapsto \lag - \case12\Trr{\hol^{-1} \dot\hol \, \dtgen },
\eeq
and this is not a total time derivative. So, the ordinary translations are \emph{not} symmetries of the action. One can also easily see that the Poisson bracket is not invariant under ordinary translations. But the deformed translations \eref{translation} are canonical transformations, and the Lagrangian is invariant. Only in the low momentum limit, where we can neglect the last term in \eref{translation} and set $\moms\approx1$, the deformed translations tend to the usual ones.

Let us close the classical part with this somewhat peculiar observation. The model that we derived is, for large momenta, not invariant under ordinary translations, although the position of the particle lives in flat Minkowski space. There is however, and maybe this is even more remarkable, still a Poincar\'e group of canonical transformations acting on the phase space, which is generated by the components of the momentum and angular momentum. It is thus not the isometry group itself that is ``deformed'', but only its action on the phase space variables. Under a translation, the particle is not shifted into the direction of the translation, but it receives an additional shift into a direction determined by its momentum. At the quantum level, this absence of the ordinary translational symmetry will show up even more drastically.

\section{Quantum theory} 

We now want to go over to a quantum description of the point particle, based on the classical reduced phase space formulation. We shall first consider the quantization of the phase space $\phsp$ itself, without imposing the mass shell condition. The philosophy behind this is that we would first like to study the quantum spacetime on which the particle lives. Typical questions that we would like to answer are:
\begin{itemize}
\item[-] What is the spectrum of the operators $\xposv_a$, $\momv_a$, and $\angv_a$?
\item[-] How are the uncertainty relations modified? 
\item[-] Does a shortest length scale show up in the model?
\end{itemize}
Later we will consider the dynamics of the particle by imposing the mass shell constraint, which translates into a kind of Klein Gordon equation, or even a Schr\"odinger like time evolution equation.

We have seen in the previous sections that the phase space $\phsp$ is the cotangent bundle $\tang_*\grpSL(2)$, or $\grpSL(2)\times\algsl(2)$. A point in that space was represented by a pair $(\hol,\xpos)$, where $\hol\in\grpSL(2)$ is the momentum and $\xpos\in\algsl(2)$ the position of the particle. The most remarkable feature of the Poisson structure on $\phsp$ was that the components of $\xpos$, and hence the coordinates of the particle, do not commute. This implies that we cannot represent quantum states as wave functions in spacetime. We can however set up a momentum representation in which the wave function depends on $\hol$.

\subsec{Momentum representation}

From the mathematical point of view, this is the most natural way to quantize the given phase space, which is the cotangent bundle of the momentum space. Hence, we take as our Hilbert space $\hlsp = \mathcal{L}_2(\grpSL(2),\d\hol)$, where $\d\hol$ is the Haar measure. A state $\ket\Psi\in\hlsp$ is then represented by a wave function $\Psi(\hol)$. Using the bra-ket notation, we can write $\Psi(\hol) = \braket{\hol}{\Psi}$, where $\ket{\hol}$ are the momentum eigenstates, $\opr{\hol}\ket{\hol} = \hol\ket{\hol}$. They are orthonormal and complete,
\beq
  \Braket{\hol_1}{\hol_2} =  \delta(\hol_1\inv\hol_2), \qquad
    \int\d\hol \, \Ket{\hol}\Bra{\hol} = \opr1.
\eeq
Using this, we can write the scalar product and the expansion in terms of a wave function as
\beq[state-expand]
  \Braket\Psi\Phi =  \int \d\hol \, \Psi^*(\hol) \, \Phi(\hol),
  \qquad
  \Ket\Psi =  \int \d\hol \, \Psi(\hol) \,  \Ket{\hol}.
\eeq
To define an operator representation for phase space functions, we demand that real functions should be represented by Hermitian operators, and that for the basic variables we have
\beq
     \pois AB = C \follows \comm AB = -\i\hbar\, \opr C.
\eeq
In particular, we want that 
\beq
     \pois {\xposv_a}\hol = - \hol \gam_a  
     \follows \comm{\xposv_a}{\hol} = \i\hbar\, \opr\hol \gam_a.
\eeq
From this we infer that $\xposv_a$ has to act on the basis states as
\beq
  \opr\xposv_a  \Ket{\hol}  =  - 2\i\hbar \, \Lvec_a  \Ket{\hol},
\eeq
where the $\Lvec^a$ are the left invariant vector fields acting on $\hol$, as defined in \eref{LR-fields}. The Haar measure is invariant under the action of the left and right invariant vector fields, so that the operator $\xposv_a$ is Hermitian. Similarly, we can derive the operator for the angular momentum from
\beq
     \pois {\angv_a}\hol = \case12[\hol,\gam_a]   
     \follows \comm{\angv_a}{\hol} = 
       -\case12 \i\hbar\, (\opr\hol\gam_a - \gam_a\opr\hol),
\eeq
which gives
\beq
  \opr\angv_a  \Ket{\hol}  =  \i\hbar \, (\Lvec_a+\Rvec_a)  \Ket {\hol}.
\eeq
Clearly, this is again Hermitian. We can express it in terms of $\xposv^a$ and $\hol$ as
\beq
   \i\hbar \,(\Lvec_a + \Rvec_a) \Ket{\hol} = 
   \i\hbar \,(\Lvec_a - \holv_a\^b\Lvec_b)  \Ket{\hol} = 
    \case12 (\opr\holv_a\^b \, \opr\xposv_b  - \opr\xposv_a )  \Ket{\hol},
\eeq
which is the same as the classical expression \eref{Ja}. There are no ordering ambiguities, because $[\opr\holv_a\^b,\opr\xposv_b]=0$.

\subsec{Euler angle representation}

To derive the spectrum of the operators $\xposv_a$, $\momv_a$, and $\angv_a$, we have to make things a bit more explicit. For this purpose, we switch to a slightly different representation. Actually, we only change the naming of the basis states. Instead of using the matrix $\hol$ to label the state $\ket\hol$, we use the Euler angles as coordinates on the group manifold. The state $\ket{\eurh,\euch,\euph}$ is then the same as $\ker{\hol}$, where
\beq
   \hol =  \expo{\frac12(\eurh+\euph)\gam_0}
         \, \expo{\euch \, \gam_1 }
         \, \expo{\frac12(\eurh-\euph)\gam_0}.
\eeq
The periodicity of the coordinates can then be expressed in the relations
\beq[rh-ph-per]
  \Ket{\eurh,\euch,\euph} = \Ket{\eurh+2\pi,\euch,\euph} = \Ket{\eurh,\euch,\euph+2\pi}  .
\eeq
To write down the scalar product and the wave function expansion, we need the Haar measure in terms of these coordinates, which is given by
\beq
   \d\hol = \frac{1}{2\pi^2} \, \sinh(2\euch) \, \d\eurh \, \d\euch \, \d\euph
\eeq
Hence, we have 
\beq[euler-state]
 \Ket\Psi 
     =  \frac{1}{2\pi^2} \int \sinh(2\euch) \, \d\eurh \, \d\euch \, \d\euph \,
                 \Psi(\eurh,\euch,\euph) \Ket{\eurh,\euch,\euph},
\eeq
and the scalar products of two such states is given by 
\beq[euler-prod]
 \Braket\Phi\Psi 
    =  \frac{1}{2\pi^2} \int \sinh(2\euch) \, \d\eurh \, \d\euch \, \d\euph \,
                 \Phi^*(\eurh,\euch,\euph) \, \Psi(\eurh,\euch,\euph).
\eeq
Of course, the states $\ket{\eurh,\euch,\euph}$ are still momentum eigenstates, and the eigenvalues are given by 
\beq
       \opr\moms   \aln\to   \cosh\euch \, \cos\eurh , \qquad
       \opr\momv_1     \to   \sinh\euch \, \cos\euph , \nwl
       \opr\momv^0 \aln\to   \cosh\euch \, \sin\eurh , \qquad
       \opr\momv_2     \to   \sinh\euch \, \sin\euph .
\eeq

\subsec{Time and angular momentum eigenstates}

Now, let us consider the $0$-components of the position and the angular momentum operator. Physically, these correspond to the time $\TTT=\xposv^0$ and the spatial angular momentum $\MMM=\angv^0$, which is the momentum associated with spatial rotations. To find their action on the basis states, we can use that
\beq
  \frac{\del \hol}{\del\eurh}  = 
    ( \Rvec^0  -  \Lvec^0 ) \, \hol , \qquad 
  \frac{\del \hol}{\del\euph}  = 
    ( \Rvec^0  +  \Lvec^0 )  \, \hol .
\eeq
From this we infer that 
\beq  
 \fl  \opr\TTT \Ket{\eurh,\euch,\euph} = 
     \i\hbar \, \Big( \deldel\eurh - \deldel\euph \Big)\, 
                                  \Ket{\eurh,\euch,\euph},  \qquad
      \opr\MMM \Ket{\eurh,\euch,\euph} = 
     \i\hbar \, \deldel\euph \,   \Ket{\eurh,\euch,\euph}.
\eeq
The eigenstates of these operators are then easily found to be 
\beq[t-m-eigenstates]
 \fl  \Ket{\ttt,\mmm;\psi} = 
   \frac{1}{2\pi^2} \int \sinh(2\euch) \, \d\eurh \, \d\euch \, \d\euph \,
   \expo{\i(\ttt+\mmm)\eurh} \, \expo{\i\mmm\euph} \, \psi(\euch) \, 
                                        \Ket{\eurh,\euch,\euph},
\eeq
where $\psi$ is an arbitrary function that depends on $\euch$ only. The eigenvalues are
\beq[t-m-eigenvalues]
  \opr \TTT \Ket{\ttt,\mmm;\psi} =  \ttt \hbar \Ket{\ttt,\mmm;\psi}, \qquad
  \opr \MMM \Ket{\ttt,\mmm;\psi} =  \mmm \hbar \Ket{\ttt,\mmm;\psi}.
\eeq
Both quantum numbers $\ttt$ and $\mmm$ have to be integers, because both $\eurh$ and $\euph$ have a period of $2\pi$. For the spatial angular momentum $\MMM$, this is of course what we should expect. It is quantized in units of $\hbar$. The quantization of the eigenvalues of $\TTT$ is however more interesting. It implies that time is quantized. It only takes values which are integer multiples of $\hbar$. In our units with $\newton=1/4\pi$, $\hbar$ also has the dimension of a length or time. We define this to be the Planck length, $\lpl=4\pi\newton\hbar$, and use the symbol $\lpl$ in the following instead of $\hbar$ whenever it appears as a length or time scale.

Besides the discreteness, there is another peculiar feature of the spectrum of $\TTT$. It starts at $\ttt=0$, and not, say, at $\ttt=\case12$ or $\ttt=0.62$. What is so special about the integer values of $\ttt$?  The answer is that there is actually nothing special. The origin of the spectrum is due to an ambiguity in the definition of the Hilbert space, or rather the operator representation introduced above, which we ignored so far. The classical phase space is not simply connected. As a result, there is an phase ambiguity in the definition of the basis states $\ket{\eurh,\euch,\euph}$. They have to be periodic in $\eurh$. However, instead of \eref{rh-ph-per}, it is sufficient ro require
\beq
  \Ket{\eurh+2\pi,\euch,\euph} = \expo{2\pi\i\tau} \Ket{\eurh,\euch,\euph}, 
\eeq
for some real number $\tau$. Doing so, it is not difficult to see that the spectrum of $\TTT$ is shifted and becomes $(\tau+\Zset)\lpl$, instead of $\Zset\,\lpl$. We can shift it wherever we like by choosing a suitable ``quantization parameter'' $\tau$. But the time steps will always have the same size $\lpl$. For simplicity, we shall restrict to $\tau=0$. Although the details of the ``spectrum of spacetime'' to be derived in the following depend on $\tau$, the \emph{qualitative} behaviour will always be the same, and this is what we are mainly interested in.

\subsec{The spectrum of spacetime}

To derive something like a position representation for the particle, we should try to diagonalize as many $\xposv$-variables as possible. We cannot diagonalize all three components, because they do not commute. But we can at least achieve to find eigenstates of two functions thereof. In fact, there is an obvious second function that commutes with $\TTT=\xposv^0$, and can be diagonalized simultaneously. As the position operators obey an algebra that is similar to the Euclidean angular momentum algebra, it is not surprising that this is the ``Casimir'' $\SSS=\xposv^a\opr\xposv_a$. It is invariant under Lorentz transformations, so it also commutes with $\MMM=\angv^0$, and we can stick to the quantum numbers $\ttt$ and $\mmm$. On the $\hol$-eigenstates, the operator $\SSS$ acts as
\beq
  \opr\SSS \Ket\hol = - 4\,\lpl^2 \, \Lvec_a \Lvec^a \Ket\hol 
                    = - 4\,\lpl^2 \, \laplace \Ket\hol ,
\eeq
where $\laplace$ is the Laplace Beltrami operator, which is equal to the square of the left (or right) invariant vector fields. This leads us to the harmonic analysis on $\grpSL(2)$. Without going too deep into the mathematical details, we give a brief derivation of the basic results in \ref{harmonic}. They are summarized as follows. The eigenstates can be split into two classes, the continuous and the discrete series. For the continuous series, we have states labeled by a positive real quantum number $\scon$, and the two integers $\ttt$ and $\mmm$, such that
\beq[cont]
   \opr\SSS \Ket{\ttt,\mmm,\scon} 
     =  (\scon^2+1) \, \lpl^2 \Ket{\ttt,\mmm,\scon}.
\eeq
The discrete series is labeled by another integer $\sdis$, and the eigenvalues are given by
\beq[disc]
   \opr\SSS \Ket{\ttt,\mmm,\sdis} 
       = - \sdis ( \sdis - 2 ) \, \lpl^2 \Ket{\ttt,\mmm,\sdis}.
\eeq
Here, the range of the quantum numbers $\ttt$, $\mmm$ and $\sdis$ are restricted as follows. For a fixed $\ttt$, the positive integer $\sdis$ only takes finitely many values, namely
\beq
    2 \le \sdis \le |\ttt| , \qquad \sdis\equiv\ttt \, (\mod 2).
\eeq
Hence, for even $\ttt$ we have $\sdis=2,4,6,\dots,|\ttt|$, and $\sdis=3,5,7,\dots,|\ttt|$ for odd $\ttt$. There are no discrete states for $\ttt$ equal to $-1$, $0$, or $1$. The values for $\mmm$ are also restricted. For given $\ttt$ and $\sdis$, we must have
\beq 
      \mmm \ge -\case12(\ttt-\sdis) \txt{for} \ttt\ge2 , 
      \qquad \mmm \le \case12(\ttt-\sdis) \txt{for} \ttt\le-2 . 
\eeq
Now, observe that the continuous series has positive eigenvalues of $\SSS$, starting at $\SSS_\min=\lpl^2$. This corresponds to \emph{spacelike} vectors $\xposv^a$. The discrete series has zero and negative eigenvalues and corresponds to \emph{lightlike} or \emph{timelike} vectors $\xposv^a$. We can illustrate the situation as follows. In a commutative spacetime, we could set up a position representation of the Hilbert space, in which each point in spacetime corresponds to one basis state. Here, we can no longer assign a basis state to every point in spacetime. We can only fix the values of $\SSS=\xposv^a\xposv_a$ and $\TTT=\xposv^0$. Alternatively, we may also use the quantity $\RRR = (\xposv_1)^2 + (\xposv_2)^2 = \SSS+\TTT^2$ instead of $\SSS$. Its spectrum is strictly positive,
\beq[R-spec]
  \opr\RRR \Ket{\ttt,\mmm,\scon} 
        \aln= (\ttt^2+\scon^2+1)\,\lpl^2 \Ket{\ttt,\mmm,\scon} ,\nwl
  \opr\RRR \Ket{\ttt,\mmm,\sdis} 
        \aln= (\ttt^2-\sdis(\sdis-2))\,\lpl^2 \Ket{\ttt,\mmm,\sdis},
\eeq
and a fixed value of $\RRR$ defines a \emph{circle} in space. Hence, we cannot associate states with points in spacetime, but at least with circles, labeled by the time coordinate $\TTT$ and the spatial radius squared $\RRR$. This reflects the fact that a state cannot be localized in spacetime, due to the commutator relation and thus the uncertainty relation between the coordinates.
\begin{figure}[t]
\begin{center}
\epsfbox{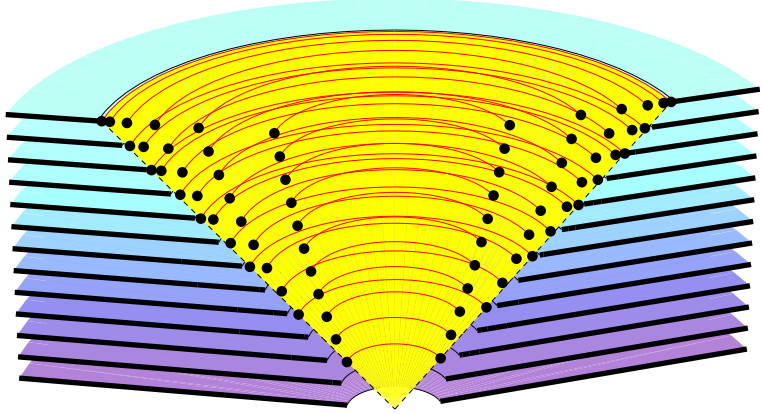}
\caption{The spectrum of quantum spacetime. Each circle in this diagram corresponds to a tower of states labeled by the spatial angular momentum quantum number $\mmm$. The position of the circles in Minkowski space are determined by the quantum numbers $\ttt$ and $\sss$.}
\label{qst}
\end{center}
\hrule
\end{figure}

A picture of this quantum spacetime is shown in \fref{qst}. It shows a circle for each allowed value of $\RRR$ and $\TTT$, for $0\le\TTT\le12\lpl$. We get discrete layers of states, corresponding to the discrete eigenvalues of $\TTT$. Outside the light cone, we have the continuous spectrum of $\RRR$, starting at $\RRR_\min=\TTT^2+\lpl^2$, a little bit off the light cone. On the light cone, we have the discrete lightlike states for $|\TTT|\ge2\lpl$, and inside the light cone the discrete timelike states which exist for $|\TTT|\ge3\lpl$. Each ring in this picture corresponds to an infinite tower of states, labeled by the angular momentum quantum number $\mmm$. In the spacelike region, where the continuous set of rings forms a plane, any value of $\mmm$ is allowed. On the light cone and in the timelike region, $\mmm$ has a lower limit. On the innermost rings, where $\sdis=\ttt$, $\mmm$ must be positive or zero. The further we go away from the time axis towards the light cone, the more negative values are allowed for $\mmm$. In the past light cone, the situation is just the reverse. There, only negative $\mmm$ are allowed on the innermost rings, and more and more positive ones are mixed in if we move towards the light cone.

This semi discrete structure of spacetime is a consequence of the fact that the coordinates $\xposv^a$ do not commute. In this sense, \fref{qst} can be thought of as a picture of a non-commutative spacetime. There is also a new kind of uncertainty relation in the theory. It is impossible to determine any two of the three $\xposv^a$ simultaneously with infinite precision. For example, we have
\beq
  \comm{\xposv_1}{\xposv_2} = 2 \i\lpl \, \opr\xposv^0 = 2\i\lpl \, \opr\TTT 
            \follows
    \expect{\Delta\xposv_1} \, \expect{\Delta\xposv_2} 
    \ge 2 \lpl \, \expect{\TTT}.
\eeq
This tells us that it is not possible to find a state that is localized inside a spacelike circle with a radius smaller than $\sqrt{2\lpl\TTT}$. This is exactly the behaviour of the innermost rings in the timelike region of \fref{qst}. The radius of the ring that is closest to the time axis, and therefore represents the ``most localized'' state at the time $\TTT$, grows with the square root of $\TTT$. Note, however, that the only completely localized state that would be allowed by the uncertainty relation above, namely at the ``origin of spacetime'' $\xpos=0$, does not exist. For low values of $\TTT$ the uncertainty is therefore even stronger.

\subsec{The origin and orientation of spacetime}

We already saw at the classical level that the theory is not invariant under ordinary translations $\xpos\mapsto\xpos+\dtgen$. At the quantum level, this can be seen in the spectrum of spacetime, which is far from being invariant under translations. The origin seems to play a special role, in the sense that it is the light cone emerging from \emph{this} point that splits the spectrum into the discrete and the continuous series. The picture is also not invariant under Lorentz transformations. However, this is only due to our choice of a special time axis and not of a physical nature. We have the same situation for the angular momentum eigenstates in three dimensional Euclidean space. There one also has to pick out a special $z$-axis to introduce the $l$ and $m$ quantum numbers, and this brakes the rotational invariance. However, the origin of angular momentum space is a well defined physical concept. It corresponds to a system that does not rotate. But what is the physical meaning of the special point in spacetime?

First of all, we should note that this special point is a \emph{parameter} that enters the action. Implicitly, it shows up in the definition of the boundary term \eref{lag-bnd}, which we added to the Lagrangian to render it finite. This term was actually determined up to a constant only, that is, up to a term that does not depend on the regularization parameter $\rlim$. There is not much choice for such a term, if we demand that it is linear in the velocities, that the Lorentz symmetry is not broken, and that the equations of motion are not modified. What we then still can do is to replace the embedding field $\fff$ by $\fff-\rtpar$. It results in a redefinition of the Lagrangian,
\beq[lag-shift-2]
    \lag \mapsto \lag + \case12 \Trr{\hol^{-1}\dot\hol\,\rtpar}.
\eeq
This does \emph{not} change the classical equations of motion. It \emph{does} however change the Poisson structure on the phase space and thus the algebra of the quantum operators. The consequence is that the origin of the spectrum shown in the figure is shifted to the point $\rtpar$. We can also see that the transformation \eref{lag-shift-2} exactly cancels \eref{lag-shift-1}, which resulted from an ordinary translation of the position coordinates. The ordinary translations are, in a sense, represented as a kind of ``metasymmetry''. The Lagrangian itself is not invariant, but it contains a parameter, and changing this parameter compensates for the translation. We should view upon this as a parameter of the theory, to be determined by ``experiments''.

So, we can shift this origin of spacetime wherever we like, but we cannot avoid that there is such a special point. On the other hand, we saw that, even with the origin fixed, there is a perfectly well defined action of the Poincar\'e group on the classical phase space $\phsp$, realized by the Lorentz rotations and the deformed translations. Of course, these symmetries are also realized as unitary transformations on $\hlsp$, generated by the operators $\angv_a$ and $\momv_a$. However, the deformed translations mix the position and momentum variables in a somewhat strange way, so that the result is not a simple shift of the spectrum of spacetime. One could interpret this as a ``deformed principle of relativity''. If one observer sees a classical state $(\hol,\xpos)$, or a quantum state $\ket\Psi$, then another observer, which is located at a different point in spacetime, sees a different state, and this state is not obtained by a simple shift in $\xpos$ but by a deformed translations \eref{translation}, or its quantum analogue. This would restore the translational invariance, and it also provides a physical interpretation of the origin, which becomes the position of the observer.

Of course, all these statements are somewhat counterintuitive. A different way out of this dilemma might be to consider multi particle models. For a system of finitely many particles, it turns out that the Poisson structure on the reduced phase space is very similar to \eref{pois}, except that now the variables $\xpos$ and $\hol$ become the \emph{relative} coordinates. They represent the relative distance between two particles, and something like their relative momentum \cite{Matschull}. As a result, one finds that the spectrum of the relative coordinates looks like the one shown in \fref{qst}. The origin then has a well defined meaning. It corresponds to a state where the two particles are at the same point in spacetime. The fact that there is no such state in the quantum spectrum, and that for the state that is closest to the origin we have $\SSS_\min=\lpl^2$, implies that the particles cannot come closer to each other than one Planck length.

Another issue, which is also due to an ambiguity in the definition of the boundary term at infinity, concerns the inherent orientation of spacetime. At the classical level, this is already apparent from the basic brackets between the coordinates $\xposv^a$. The appearance of the Levi Civita symbol indicates that the theory is not ``$PT$'' invariant. The Lagrangian is not invariant with respect to the substitutions $\hol \mapsto \hol^{-1}$ (or $\mom \mapsto -\mom$) and $\xpos \mapsto- \xpos$. Again, this broken symmetry can be seen quite nicely in the quantum spacetime picture. On the discrete rings in the future light cone which are closest the time axis, only positive values of the spatial angular momentum occur. The further we go away from the time axis, the more negative angular momenta mix in. In the past light cone, the situation is reversed. Slightly sloppy speaking, if we want to localize the particle on such a ring, then we have to give it an angular momentum in a certain direction.

This orientation results from another freedom in the definition of the boundary term \eref{lag-bnd}. We can replace $\fff_+$ by a linear combination of $\fff_+$ and $\fff_-$, which leads to a more general reduced Lagrangian, whose kinetic part is parametrized by a real number $\sigma$, 
\beq
  \lagkin = - \case12\sigma\, \Trr{\dot\hol\,\hol^{-1} \xpos} 
       - \case12 (1-\sigma) \, \Trr{\hol^{-1}\dot\hol\,\xpos}.
\eeq
Again, this does not change the equations of motion, but it does affect the Poisson structure. Replacing $\sigma$ with $1-\sigma$ corresponds to a ``$PT$'' transformation $\xpos\mapsto-\xpos$, $\hol\mapsto\hol^{-1}$, so that this symmetry is also realized as a kind of metasymmetry. Unfortunately, the value $\sigma=\case12$, which would remove the handedness of spacetime, is the only value for $\sigma$ for which the classical equations of motion are no longer reproduced fully. This can be seen by repeating the derivation of the evolution equations \eref{red-evolve} with the modified Lagrangian. So, we have to take the same point of view here as well. The orientation of spacetime is a parameter which has to be determined by experiments. For simplicity, let us stick to $\sigma=0$ in the following.

\subsec{The physical Hilbert space} 

Let us now consider the dynamics of the particle on the quantum spacetime, which is described by the mass shell constraint. Classically, it was given by
\beq[const]
  \con =   \moms - \cos\mass = 0 .
\eeq
Let us see how the corresponding operator acts in the Euler angle representation. We find that
\beq
   \opr\con \Ket{\eurh,\euch,\euph}    \aln= 
  (\opr\moms - \cos \mass) \Ket{\eurh,\euch,\euph} 
   \nwl     \aln= 
  (\cos \eurh \, \cosh\euch - \cos \mass)  \Ket{\eurh,\euch,\euph}.
\eeq
If we expand a generic state as in \eref{euler-state},
\beq
  \Ket{\Psi} = 
     \frac1{2\pi^2} \int \sinh(2\euch)\,\d\eurh\,\d\euch\,\d\euph\, 
           \Psi(\eurh,\euch,\euph) \, \Ket{\eurh,\euch,\euph},
\eeq
then the wave function has to satisfy
\beq
  ( \cos\eurh \, \cosh\euch - \cos\mass ) 
                       \, \Psi(\eurh,\euch,\euph) = 0.
\eeq
It states that $\Psi$ has support on the mass shells only, and defines the physical state space $\phhs$ as a subset of $\hlsp$. We can use $\euph$ and $\euch$ as coordinates on the mass shells, and write the general solution to the mass shell constraint as
\beq[phys-state-ch-ph]
  \Psi(\eurh,\euch,\euph) = \sum_{\anti} 
    \delta( \cos \eurh \, \cosh \euch - \cos \mass) \, 
     \theta( \anti \, \sin\eurh) \, \Psi_\anti(\euch,\euph).
\eeq
Here, $\Psi_\anti(\euch,\euph)$, with $\anti=\pm1$, is a pair of arbitrary functions. We have $\anti=1$ for particles and $\anti=-1$ for antiparticles. The theta function, which is $0$ for negative arguments and $1$ for positive, ensures that $0<\eurh<\pi$ for particles and $-\pi<\eurh<0$ for antiparticles.

As for the relativistic point particle, we are faced with the problem that the physical state space $\phhs$ is actually not a subspace of the Hilbert space $\hlsp$. The wave functions \eref{phys-state-ch-ph} are not normalizable. The problem can be solved in the standard way, by introducing a new scalar product, which is, up to a constant, uniquely determined if we require it to be Lorentz invariant. In other words, we want the operators $\angv_a$ and $\momv_a$ to remain Hermitian under the new product.

To find this physical product explicitly, we note that the wave function $\Psi_\anti(\euch,\euph)$ in \eref{phys-state-ch-ph} transforms as a scalar under Lorentz transformations. The left hand side is a scalar function of $\hol=\hol(\eurh,\euch,\euph)$, and the argument of the delta function is also a scalar, because it is equal to $\moms-\cos\mass$. Hence, to define a Lorentz invariant scalar product, we need a Lorentz invariant measure on the mass shell in terms of the coordinates $\euch$ and $\euph$. This is found to be
\beq[phys-prod]
 \pBraket\Phi\Psi   
    = \frac1{4\pi^2} \sum_\anti 
            \int \frac{2 \sinh(2\euch) \, \d\euch\, \d\euph}
                      {\sqrt{\sinh^2\euch + \sin^2\mass}} \, 
     \Phi^*_\anti(\euch,\euph) \,  \Psi_\anti(\euch,\euph) .
\eeq
It becomes a more familiar expression, if we replace the polar coordinates $(\euch,\euph)$ by the components of the spatial momentum vector $\smom = (\momv_1,\momv_2)$. Explicitly, we have
\beq[qq-ph-rh]
  \momv_1 = \sinh\euch \, \cos\euph ,\qquad
  \momv_2 = \sinh\euch \, \sin\euph ,
\eeq
and this implies
\beq[q-measure]
 \frac{2 \sinh(2\euch) \, \d\euch\, \d\euph}
      {\sqrt{\sinh^2\euch + \sin^2\mass}}
            = \frac{\d^2\smom}{\sqrt{\smom^2+\sin^2\mass}} 
            = \frac{\d^2\smom}{\omega(\smom)}.
\eeq
On the mass shell, $\omega(\smom)$ is equal to $|\momv^0|$, so that the expression for the scalar product becomes formally the same as for the relativistic point particle in the momentum representation,
\beq
  \pBraket\Phi\Psi 
  = \frac1{4\pi^2} \sum_\anti \int \frac{\d^2\smom}{\omega(\smom)}
   \, \Phi^*_\anti(\smom) \Psi_\anti(\smom).
\eeq

\subsec{The Klein Gordon equation}

As a next step, we would like to see in which sense the mass shell constraint can be interpreted as a time evolution equation for the quantum state. To explain what we are going to do, let us consider a relativistic point particle in flat spacetime for a moment. There we have the mass shell constraint in momentum space,
\beq
    ( \momv_a \momv^a + \mass^2) \, \Psi(\mom) = 0.
\eeq
Clearly, to see that this is actually a time evolution equation, we only need to Fourier transform to the position representation. But what if there was no such representation, as in our model?  We can then still see that it is a time evolution equation, if we diagonalize another set of operators, such that one of them is the time operator $\TTT=\xposv^0$. Let us choose as the other two operators the absolute length of the spatial momentum $\momv^2 = \momv_1^2+\momv_2^2$, and the spatial angular momentum $\MMM = \xposv_1\momv_2 - \xposv_2\momv_1$. They commute with each other and both commute with $T$. If we now write the transformed wave function as $\Psi(\ttt,\mmm;\momv)$, where $\ttt\in\Rset$, $\mmm\in\Zset$, and $\momv\in\Rset_+$, then the mass shell constraint becomes a second order differential equation in $\ttt$,
\beq[KG]
  - \hbar^2 \frac{\del^2}{\del \ttt^2} \, \Psi(\ttt,\mmm;\momv) 
    = (\momv^2+\mass^2) \, \Psi(\ttt,\mmm;\momv).
\eeq
So, without transforming completely to the position representation, we find this ``Klein Gordon equation''. We can convert it into a Schr\"odinger equation, if we decompose the solutions into particle and antiparticle states. The general solution to \eref{KG} can then be written as
\beq
   \Psi(\ttt,\mmm;\momv) = \sum_{\anti} \Psi_\anti(\ttt,\mmm;\momv), 
\eeq
where the two functions $\Psi_\anti$, with $\anti=\pm1$, have to obey
\beq
  - \i \hbar \deldel t \, \Psi_\anti(\ttt,\mmm;\momv) = 
        \anti\, \sqrt{\momv^2+\mass^2}\, \Psi_\anti(\ttt,\mmm;\momv).
\eeq
We chose this somewhat skew representation, because in our model we can diagonalize exactly the same operators. We did this already in \eref{t-m-eigenstates}, which was the most general eigenstate of $\TTT$ and $\MMM$. Let us define
\beq
  \Ket{\ttt,\mmm;\euch} = 
   \frac{1}{4\pi^2} \int  \d\eurh \, \d\euph \,
   \expo{\i(\ttt+\mmm)\eurh} \, \expo{\i\mmm\euph}  \, \Ket{\eurh,\euch,\euph},
\eeq
and expand the state in this basis, 
\beq
  \Ket\Psi = \sum_{\ttt\in\Zset} \sum_{\mmm\in\Zset} 
           \int \d\euch \, 2 \sinh(2\euch) \,
              \Psi(\ttt,\mmm;\euch) \, \Ket{\ttt,\mmm;\euch}.
\eeq
As the rapidity $\euch$ is related to the spatial momentum by $\momv=\sinh\euch$, this is essentially the same representation as the one considered above for the relativistic point particle. To write the mass shell constraint as an equation for the wave function $\Psi(\ttt,\mmm;\euch)$, we need to know how $\moms$ acts on the new basis states,
\beq
   \opr\moms \Ket{\ttt,\mmm;\euch}     \aln= 
   \frac{1}{4\pi^2} \int  \d\eurh \, \d\euph \,
   \expo{\i(\ttt+\mmm)\eurh} \, \expo{\i\mmm\euph}  \, 
       \cos\eurh \, \cosh\euch \Ket{\eurh,\euch,\euph} \nwl
       \aln= \case12 \cosh\euch \, \big(
       \Ket{\ttt+1,\mmm;\euch} + \Ket{\ttt-1,\mmm;\euch} \big).
\eeq
Using this, we can write the Klein Gordon equation as 
\beq[KGdiff]
  \Psi(\ttt+1,\mmm;\euch) + \Psi(\ttt-1,\mmm;\euch) = 
     2 \frac{\cos\mass}{\cosh\euch} \, \Psi(\ttt,\mmm;\euch).
\eeq
This is not a second order \emph{differential} equation, but a second order \emph{difference} equation. We should have expected something like this, because the eigenvalues of $\TTT$ are discrete. The fact that it is such a nice nearest neighbour equation is a result of the appearance of $\cos\eurh$ in the constraint. It is not difficult to see that it reduces to the ordinary Klein Gordon equation in the limit $\lpl\to0$. To do this, we set $\sinh\euch=\momv$, and then make the replacements
\beq
  \mass \mapsto 4\pi\newton \, \mass, \qquad
  \momv \mapsto 4\pi\newton \, \momv, \qquad
  \ttt \mapsto \ttt / \lpl,
\eeq
to restores the physical units. If we then expand \eref{KGdiff} for small $\lpl$, but finite $\hbar$, and use that $\lpl = 4\pi\newton\hbar$, we recover \eref{KG} in the leading order, which is $\lpl^2$.

We can also convert the discretized Klein Gordon equation into a Schr\"odinger equation, if we split the wave function into a particle and an antiparticle wave function. This becomes more or less straightforward, if we first write the step equation as
\beq
  \Psi(\ttt+1,\mmm;\euch) + \Psi(\ttt-1,\mmm;\euch) 
      = (\unit + \unit^{-1} ) \Psi(\ttt,\mmm;\euch),
\eeq
where $\unit$ is found to be 
\beq[unit]
  \unit = \frac{\cos\mass + \i\sqrt{\sinh^2\euch+\sin^2\mass}}
                   {\cosh\euch}.
\eeq
This is a complex unit, so it acts on wave functions as a unitary operator. We can now write the general solution to the second order difference equation as a sum of solutions to the following first order difference equations,
\beq[time-p-ap]
  \Psi(\ttt,\mmm;\euch) = \sum_\anti \Psi_\anti(\ttt,\mmm;\euch) ,  
    \quad \Psi_\anti(\ttt+1,\mmm;\euch) 
               = \unit^\anti \, \Psi_\anti(\ttt,\mmm;\euch).
\eeq
This is very similar to the time evolution equation in ordinary quantum mechanics, 
\beq
    -\i\hbar \deldel t \Psi(t) = \ham \,\Psi(t)   \equivalent
   \Psi(t+\tau) = \expo{\i\ham\tau/\hbar} \Psi(t).
\eeq
The only difference between our time evolution equation and the Schr\"odinger equation is that we cannot write it as a differential equation, because we do not have a continuous time. We can however use it to read off what the energy of our particle is. For this purpose, we have to write the unitary operator $\unit^\anti$, which moves us forward one unit in time, as the exponential of an Hermitian operator,
\beq
  \unit^\eps = \expo{\i\ham}
             = \cos\ham + \i\sin\ham .
\eeq
If we compare this to \eref{unit}, we find that $\ham$ is determined by 
\beq
  \cos\ham = \frac{\cos\mass}{\cosh\euch} ,\qquad
  \anti\sin\ham > 0 .
\eeq
These are exactly the equations that determine the value of the Euler angle $\eurh$ on the mass shell. So we have, on the mass shell, $\ham=\eurh$, and we recover, at the quantum level, exactly what we already found at the classical level in the end of \sref{particle}. The energy of the particle is given by the Euler angle $\eurh$, which is related to the deficit angle of the particle, measured in the rest frame of the observer. A peculiar feature of this quantum energy is however that it is only determined up to a multiple of $2\pi$ (times the Planck energy), because $\eurh$ is an angular variable. Clearly, this is because, at the quantum level, the energy of the particle is actually its \emph{frequency}, and if time is discrete, than this can only be determined up to a multiple of $2\pi$, divided by the time step. 

\subsec{Observables}

Finally, we would like to discuss the question how to extract physical information about the particle from the state $\ket\Psi$. For this purpose, we need a complete set of \emph{observables}, that is, operators that commute with the mass shell constraint, and for which we have a physical interpretation of the corresponding classical phase space function. This is quite obvious for the operators $\momv_a$ and $\angv_a$, representing the momentum and angular momentum of the particle. They are in fact complete, but not independent. The momentum obeys the mass shell condition, $\momv_a\momv^a+\sin^2\mass=0$, and for the angular momentum we have $\angv_a\momv^a=0$. As a set of independent observables we could choose the spatial components $\momv_i$ and $\angv_i$ ($i=1,2$), but there is another set which has a more natural physical interpretation. 

The position operators $\xposv^a$ are not observables as they do not commute with the constraint. But we can use them to construct an observable which represents the position of the particle at a fixed time $\tau$. Classically, this is given by $\dposv_i(\tau) = \xposv_i - \momv_i/\momv_0 (\xposv_0+\tau)$, which is easily found to have a vanishing bracket with the constraint. To make it a Hermitian operator, we choose the symmetrized ordering
\beq
   \opr\dposv_i(\tau)     \aln= \opr\xposv_i 
   -  (\opr\xposv_0 + \tau) \, \frac{\opr\momv_i}{2\,\opr\momv_0} 
   -  \frac{\opr\momv_i}{2\,\opr\momv_0} \, (\opr\xposv_0 + \tau)  
  \nwl     \aln= \opr\xposv_i  
           - \frac{\opr\momv_i}{\opr\momv_0} \, (\opr\xposv_0 + \tau)
   + \i\hbar \,  
    \frac{\opr\momv_i \opr\moms + \eps_{ij} \opr\momv_j \opr\momv_0} 
         {2\,\opr\momv_0^{\,2}} 
\eeq
This looks again similar to the analogue of the relativistic point particle, except the the usual ``Newton Wigner term'' $\case12\i\hbar \, \opr\momv_i / \opr\momv_0^2$ is replaced by a more complicated expression. However, in the low momentum limit, it reduces to the usual one, because we can neglect $\momv_0$ compared to $\moms\approx1$.

Now, the operators $\dposv_i(\tau)$, for fixed $\tau$, and the spatial components of the momentum $\momv_i$ form a complete set of independent observables. Classically, they fix the initial condition for the particle at time $\tau$. To set up a position representation, we should diagonalize the observables $\dposv_i(\tau)$. Unfortunately, this is still not possible, because they do not commute. If we consider the operators $\dposv_i(\tau)$ as representing the position of the particle at time $\tau$, then the space on which the particle is moving is non-commutative, like to the non-commutative spacetime that we encountered before we imposed the mass shell condition. To see the similarity, let us compute the commutator of the $\dposv_i$ operators and consider the resulting uncertainty relation. For the classical phase space functions $\dposv_i(\tau)$, we find
\beq
  \pois{\dposv_i(\tau)}{\dposv_j(\tau)}  = 
    2\,\eps_{ij} \,
       \big( \tau + \frac{\momv_k \, \dposv_k(\tau)}{\momv_0} \big) .
\eeq
Hence, assuming that the same holds up to higher order corrections for the quantum operators, we find that the minimal uncertainty in the position coordinates of a particle centered at the origin grows with the square root of $\tau$. This is exactly the same behaviour as that of the innermost rings in \fref{qst}. So, we can reproduce the same result at the level of observables, that is, with the gauge degrees of freedom completely divided out.

What we cannot reproduce at this level is the discretization of time. This is because there is no observable that corresponds to the measurement of time. Note, for example, that the observable $\dposv_i(\tau)$ is constructed by a kind of interpolation, and it is defined for every real value of $\tau$. The integers do not play any special role. To reproduce the discreteness of time, one should introduce an additional degree of freedom, for example a clock associated with the particle, and then consider the eigenvalues of the operator corresponding to the measurement ``read off the clock at external time $\tau$''. Such an operator is expected to have a discrete spectrum, because it is ``canonically conjugate'' to the mass of the particle, which is restricted to a finite interval. 

\newpage

\section*{Acknowledgements} 

We would like to thank 
\begin{center}
Ingemar Bengtsson, Gerard 't~Hooft, Jorma Louko and Erik Verlinde
\end{center}
for helpful questions and interesting discussions, and also the referees for helping us to improve this article. 

\setcounter{section}{0}
\def\thesection{Appendix \Alph{section}}
\def\theequation{\Alph{section}.\arabic{equation}}
\section{The Lorentz group} 
\label{lorentz}
 
In this appendix, some basic properties of the Lorentz algebra and its associated Lie groups are collected. The notations is as follows. A vector in three dimensional Minkowski space $\Mset^3$ takes a small latin index $a,b,c,\dots$, running from $0$ to $2$. The metric is $\eta_{ab}=\diag(-1,1,1)$, and the Levi Civita symbol $\eps^{abc}$ obeys $\eps^{012}=-\eps_{012}=1$.

\subsec{The vector and spinor representation} 

The group $\grpSO_+(1,2)$ of time oriented proper Lorentz transformations acts on a vector in $\Mset^3$ from the right as
\beq[vrep]
  \avecv^a \mapsto \avecv^b \, \holv_b\^a , \txt{where}
   \holv_a\^b \holv_c\^d \eta_{bd} = \eta_{ac} , \quad
   \holv_0\^0 > 0.
\eeq
Elements of the associated Lie algebra $\algso(1,2)$ are represented by matrices $\bvecv_a\^b$ with $\bvecv_{ab}=-\bvecv_{ba}$. As a vector space, the algebra is isomorphic to $\Mset^3$ itself, and its spinor representation is $\algsl(2)$. We denote elements of $\algsl(2)$, that is, traceless $2\times2$-matrices, by bold letters. Due to the isomorphism between the $\algsl(2)$ and $\Mset^3$, we can also interpret them as vectors in Minkowski space. Explicitly, the isomorphism is given by expansion in terms of gamma matrices,
\beq[M3-sl]
   \avec = \avecv^a \, \gam_a \equivalent
   \avecv^a = \case12 \Trr{\avec \gam^a},
\eeq
where
\beq[gamma]
  \gam_0 = \pmatrix { 0 & 1 \cr -1 & 0 } , \quad
  \gam_1 = \pmatrix { 0 & 1 \cr 1  & 0 }, \quad
  \gam_2 = \pmatrix { 1 & 0 \cr 0 & -1 } .
\eeq
The algebra of these matrices reads
\beq[gamma-alg]
   \gam_a \gam_b = \eta_{ab} \, \one - \eps_{abc} \, \gam^c,
\eeq
where $\one$ is the $2\times2$ unit matrix. The scalar product on $\Mset^3 \simeq \algsl(2)$ can be written as $\avecv_a\bvecv^a = \case12\Trr{\avec\bvec}$. Timelike, lightlike and spacelike \emph{matrices} $\avec$ are defined according to the corresponding properties of the \emph{vector} $\avecv^a$. For example, $\avec$ is called positive timelike if $\Trr{\avec\avec} < 0$ and $\Trr{\avec\gam^0} > 0$. 

Elements of the group $\grpSL(2)$ are also denoted by bold letters. An element $\hol\in\grpSL(2)$ acts on a vector $\avec\in\algsl(2)$ in the adjoint representation as
\beq[srep]
  \avec \mapsto  \hol^{-1} \avec \, \hol. 
\eeq
For this to be the same as \eref{vrep}, we must have 
\beq[SL-SO]
  \hol^{-1} \gam_a \hol =  \holv_a\^b \, \gam_b \equivalent 
  \holv_a\^b = \case12 \Trr{ \hol^{-1} \gam_a  \hol \, \gam^b }.
\eeq
This is the two-to-one group homomorphism mapping the spinor representation $\grpSL(2)$ onto the vector representation $\grpSO_+(1,2)$. The corresponding isomorphism of the algebras $\algsl(2)$ and $\algso(1,2)$ is given by
\beq[sl-so]
   \avec =  - \case14 \eps^{abc} \, \avecv_{ab} \, \gam_c  
   \equivalent
   \avecv_{ab} =  \eps_{abc} \, \Trr{ \avec \gam^c}. 
\eeq
As a general convention, we always denote different representations of the same object by the same letter, but with different indices attached, or by a bold letter if it is a $2\times2$-matrix.

\subsec{The group manifold} 

As a manifold, the group $\grpSL(2)$ is naturally embedded in $\Rset^4$ (or $\Rset\times\Mset^3$). The most convenient embedding coordinates are obtained by expanding $\hol$ in terms of the unit and the gamma matrices. We define a scalar $\moms$ and a vector $\momv_a$ such that
\beq[expansion]
  \hol = \moms \, \one + \momv_a \, \gam^a  
  \follows 
  \moms^2 - \momv_a \momv^a = 1,
\eeq
where the last equations results from the condition that $\det\hol=1$. It defines a $(2,2)$-hyperboloid embedded in $\Rset^4$, which is shown in \fref{grp}. We can also define a ``projection'' of the group into the algebra,
\beq[projection]
     \hol = \moms \, \one + \momv_a \, \gam^a  
     \quad \mapsto \quad 
     \mom =  \momv_a \, \gam^a .
\eeq
The image of the group manifold under this projection is the set of all vectors $\mom\in\algsl(2)$ with length squared bigger of equal to $-1$. It is the subset of Minkowski space between the two unit timelike hyperboloids. 

A useful set of coordinates on the group manifold is provided by the ``Euler angles'' $(\eurh,\euch,\euph)$. The general solution to \eref{expansion} can be written as 
\beq[p-euler]
          \moms   \aln= \cos \eurh \, \cosh \euch , \qquad
          \momv_1     = \cos \euph \, \sinh \euch , \nwl
          \momv^0 \aln= \sin \eurh \, \cosh \euch , \qquad
          \momv_2     = \sin \euph \, \sinh \euch . 
\eeq
The two ``angular'' coordinates $\eurh$ and $\euph$ have a period of $2\pi$, and $\euch$ is a ``hyperbolic angle'' with $\euch\ge0$. For $\euch=0$, the angle $\euph$ becomes redundant, whereas $\eurh$ is always regular. All together, they form a kind of ``cylindrical'' coordinate system, with radial coordinate $\euch$, polar angle $\euph$, and $\eurh$ playing the role of the ``$z$''-coordinate, which is however winded up. The coordinates $\eurh$ and $\euch$ are shown as the grid lines in \fref{grp}. They are called Euler angles because they are closely related to the Euler angles of $\grpSU(2)$. We can write the group element corresponding to \eref{p-euler} as
\beq[euler]
   \hol = \expo{\frac12 (\eurh+\euph)\gam_0} \, 
            \expo{\euch \gam_1}
            \expo{\frac12 (\eurh-\euph)\gam_0}. 
\eeq
The Lorentz transformation represented by $\hol$ is decomposed into a rotation of the spatial plane by $\eurh+\euph$, a boost in the $\gam_2$ direction with parameter $\euch$, and another rotation by $\eurh+\euph$. To see this, and to check that \eref{euler} is actually the same as \eref{p-euler}, one has to use of the following formulas for the exponential map $\algsl(2)\to\grpSL(2)$,
\beq[exp-gam]
  \expo{\alpha\gam_0} \aln= \cos \alpha \,\, \one + \sin \alpha \,\, \gam_0, 
  \nwl
  \expo{\beta \gam_1} \aln= \cosh \beta \,\, \one + \sinh \beta \,\, \gam_1.
\eeq
They imply that the exponential of $\gam_0$ generates a rotation in the $\gam_1$-$\gam_2$ plane,
\beq[exp-rot]
  \expo{-\frac12\defangle\gam_0} \, \gam_1 
                   \, \expo{\frac12\defangle\gam_0} 
   \aln= \cos\defangle  \,\, \gam_1 + \sin\defangle \,\, \gam_2 , \nwl
  \expo{-\frac12\defangle\gam_0} \, \gam_2
                   \, \expo{\frac12\defangle\gam_0} 
   \aln=  \cos\defangle \,\, \gam_2 - \sin\defangle \,\, \gam_1 ,
\eeq
and similarly, that the exponential of $\gam_1$ generates a boost in $\gam_2$ direction and vice versa.

\subsec{Rotations and the mass shells}
 
The projection \eref{projection} can be used to distinguish between timelike, lightlike, and spacelike \emph{group elements} $\hol$, according to the properties of the vector $\mom$. The timelike ones are then those with $-1<\moms<1$. On the boundaries of this interval we have the lightlike group elements, and outside this range the spacelike ones. The vector $\mom$ points along the ``axis'' of the Lorentz transformation represented by $\hol$, which can be seen from the fact that it is invariant under the adjoint action,
\beq[hol-mom-comm]
  \hol^{-1} \mom \, \hol = \mom .
\eeq
If this axis is timelike, then the Lorentz transformation is a \emph{rotation} of the spacelike plane orthogonal to $\mom$. If it is spacelike, then it is a \emph{boost}, and in between we have the lightlike Lorentz transformations. The rotations play a special role, because they appear as the momenta of massive particles. The mass of the particle is related to the angle of rotation by $\defangle=2\mass$. To find all group elements representing a rotation by $2\mass$, consider a unit spacelike vector $\avec$ orthogonal to the axis $\mom$, so that
\beq[l-orth]
  \avec\avec = \one ,    \qquad
  \avec\mom+\mom\avec = 0  \follows  
  \hol^{-1} \avec   = \avec \hol  .
\eeq
Under the rotation, $\avec$ is mapped onto $\hol^{-1}\avec\hol$. If this is a rotation by $2\mass$, then the scalar product of this vector with $\avec$ must be $\cos(2\mass)$, independent of the choice of $\avec$. Hence, we must have
\beq
   \case12 \Trr{\avec\hol^{-1}\avec\hol}  = 
   \case12 \Trr{\avec\avec\hol\hol} = 
   \case12 \Trr{\hol\hol} = \cos(2\mass).
\eeq
If we expand $\hol$ in terms of the gamma matrices according to 
\eref{expansion}, this becomes
\beq
    \moms^2 - 1 = \cos(2\mass) 
    \equivalent  \moms^2 = \cos^2\mass.
\eeq
The solutions to this equation fall into four conjugacy classes of $\grpSL(2)$, containing the elements $\pm\expt{\pm\mass\gam_0}$. The two sign ambiguities arise as follows. The sign in the exponent fixes the \emph{direction} of the rotation. In contrast to three dimensional Euclidean space, in Minkowski space we can distinguish between rotations in positive and negative direction. The second sign appears because $\hol$ is an element of the \emph{spinor} representation of the Lorentz group, whereas ``rotation by $2\mass$'' refers to the \emph{vector} representation. Thus, for every rotation we have two spinor representations, which differ by an overall sign.

To get rid of at least one sign, we change the definition of ``rotation by an angle $2\mass$'' slightly, such that it also fixes the action on spinors. In the vector representation, we cannot distinguish a rotation in the positive direction by $2\mass>0$ from a rotation in the negative direction by $2\mass-2\p<0$. But in the spinor representation we can. The former is then represented by an element in the conjugacy class of $\expt{\mass\gam_0}$, whereas the second lies in the class of $\expt{(\mass-\pi)\gam_0}=-\expt{\mass\gam_0}$. So, if we only call the former a ``rotation by $2\mass$'', then the condition becomes
\beq[rot-con]
   \moms = \cos\mass \equivalent 
   \case12 \Trr{\hol} =  \cos\mass.
\eeq
As this equation is still invariant under $\mass\mapsto-\mass$, or $\hol\mapsto\hol^{-1}$, it still does not fix the direction of the rotation. This ambiguity remains in the description of the particle, and it becomes a kind of particle--antiparticle duality.

\section{Harmonic analysis} 
\label{harmonic}
 
Here, we shall briefly derive the basic results of the harmonic analysis on the group manifold $\grpSL(2)$. For a more detailed treatment of the subject we refer to \cite{Limic,Verlinde,Biedenharn,Vilenkin}. We start with the general $\TTT=\xposv^0$ and $\MMM=\angv^0$ eigenstates $\ket{\ttt,\mmm;\psi}$, where $\psi$ is function the Euler angle $\euch$, which is essentially the spatial momentum and commutes with $\TTT$ and $\MMM$ (see \eref{t-m-eigenstates}). The problem is to diagonalize the operator $\SSS=\xposv^a\xposv_a$ on these states, which is equal to the Laplace Beltrami operator. Explicitly, 
\beq
  \opr\SSS \Ket{\ttt,\mmm;\psi} = \Ket{\ttt,\mmm;\laplace\psi},
\eeq
where
\beq[laplace]
  \laplace = \hbar^2 
  \Big( \frac{1}{\sinh(2\euch)} \deldel\euch \sinh(2\euch) \deldel\euch
           - \frac{4\mmm}{\sinh^2(2\euch)}
           + \frac{2\mmm\ttt + \ttt^2}{\cosh^2\euch} \Big) .
\eeq
It is Hermitian with respect to the scalar porduct
\beq[t-m-prod]
  \Braket{\ttt,\mmm;\psi}{\ttt',\mmm';\phi}
   =  \delta_{\ttt,\ttt'} \, \delta_{\mmm,\mmm'} 
     \intl{0}{\infty} \d\euch \, 2 \sinh(2\euch) \, 
        \psi^*(\euch) \, \phi(\euch).
\eeq
Let us introduce the following raising and lowering operators, which change the quantum numbers $\ttt$ and $\mmm$,
\beq
  \TTT_\pm = \xposv_1 \pm \i \xposv_2 , \qquad
  \MMM_\pm = \angv_1 \mp \i \angv_2 + \case12 (\xposv_1 \mp \i \xposv_2).
\eeq
They act on $\hol$ eigenstates as 
\beq
  \opr\TTT_\pm \Ket{\hol} =  2\i\hbar \, \Lvec_\pm \Ket{\hol}, \qquad
  \opr\MMM_\pm \Ket{\hol} =  \i\hbar \, \Rvec_\mp \Ket{\hol},
\eeq
where $\Lvec_\pm=\Lvec_1\pm\i\Lvec_2$ and $\Rvec_\pm=\Rvec_1\pm\i\Rvec_2$ are combinations of the left and right invariant vector fields. Using this and $\Lvec_-\Lvec_+ = \Lvec_a\Lvec^a + \Lvec^0\Lvec^0 + \i\Lvec^0$, one finds the following relations:
\beq
  \opr\TTT_+ \opr\TTT_-   \aln=  \opr\SSS + \opr\TTT\,(\opr\TTT-2\hbar),\nwl
  \opr\TTT_- \opr\TTT_+   \aln=  \opr\SSS + \opr\TTT\,(\opr\TTT+2\hbar). 
\eeq
Similarly, with the same expressions for the right invariant vector fields, one obtains
\beq
  \opr\MMM_+ \opr\MMM_-  \aln= \case14 \opr\SSS + 
       (\opr\MMM+\case12\opr\TTT)(\opr\MMM+\case12\opr\TTT-\hbar),\nwl
  \opr\MMM_- \opr\MMM_+  \aln= \case14 \opr\SSS + 
    (\opr\MMM+\case12\opr\TTT)(\opr\MMM+\case12\opr\TTT+\hbar). 
\eeq
We also need the commutator relations 
\beq
     \comm{\TTT}{\TTT_\pm}\aln= \pm 2\hbar \, \opr\TTT_\pm , \qquad
     \comm{\TTT}{\MMM_\pm}    = 0 ,\nwl
     \comm{\MMM}{\TTT_\pm}\aln= \mp \hbar \, \opr\TTT_\pm, \qquad 
     \comm{\MMM}{\MMM_\pm}    = \pm \hbar \, \opr\MMM_\pm.
\eeq
Now, let $\ket{\ttt,\mmm,\sss}$ be a normalized eigenstate of $\TTT$, $\MMM$, and $\SSS$, with eigenvalues $\hbar\ttt$, $\hbar\mmm$, and $\hbar^2\sss$, respectively. Acting on this with the raising and lowering operators above, we can increase and decrease the quantum numbers $\ttt$ and $\mmm$,
\beq
\fl  \opr\TTT_\pm \Ket{\ttt,\mmm,\sss}   
         \propto  \Ket{\ttt \pm 2,\mmm \mp 1,\sss},    \qquad
     \opr\MMM_\pm \Ket{\ttt,\mmm,\sss}   
         \propto  \Ket{\ttt,      \mmm \pm 1,\sss}.
\eeq
If we compute the norm of these states, we find
\beq[norm]
 \fl \norm{ \opr\TTT_\pm \Ket{\ttt,\mmm,\sss}}     \aln= 
       \Bra{\ttt,\mmm,\sss} \opr\TTT_\mp \opr\TTT_\pm \Ket{\ttt,\mmm,\sss}  
 \nwl 
     \aln= \Bra{\ttt,\mmm,\sss} 
            \big( \opr\SSS + \opr\TTT(\opr\TTT \pm 2\hbar) \big) 
                                               \Ket{\ttt,\mmm,\sss} 
 \nwl 
     \aln= \hbar^2 \, (\sss + \ttt(\ttt \pm 2)). 
 \nwl
 \fl \norm{ \opr\MMM_\pm \Ket{\ttt,\mmm,\sss}}     \aln= 
        \Bra{\ttt,\mmm,\sss} \opr\MMM_\mp \opr\MMM_\pm \Ket{\ttt,\mmm,\sss}  
 \nwl 
     \aln= \Bra{\ttt,\mmm,\sss} 
          \big( \case14 \opr\SSS + (\opr\MMM+\case12\opr\TTT) 
                                   (\opr\MMM+\case12\opr\TTT\pm\hbar) \big) 
                                               \Ket{\ttt,\mmm,\sss} 
 \nwl 
     \aln= \case14 \hbar^2 \, ( \sss + (2\mmm+\ttt)(2\mmm+\ttt \pm 2)) .
\eeq
These have to be non-negative, which gives the following algebraic restrictions on the quantum numbers,
\beq[norm-cond]
\fl  \sss +  \ttt ( \ttt + 2 )   \aln\ge 0  \quad (\opr\TTT_+), \qquad 
     \sss + (2\mmm +\ttt)(2\mmm+\ttt+2) \ge 0  \quad (\opr\MMM_+), \nwl
\fl  \sss +  \ttt ( \ttt - 2 )   \aln\ge 0  \quad (\opr\TTT_-), \qquad 
     \sss + (2\mmm +\ttt)(2\mmm+\ttt-2) \ge 0  \quad (\opr\MMM_-). 
\eeq
Moreover, we infer that starting from any state $\ket{\ttt,\mmm,\sss}$, we can build up a two dimensional ``tower'' of states by acting with the raising and lowering operators on them. We can reach any value for $\mmm$ and $\ttt$, except that we cannot change the parity of $\ttt$, and provided that the series does not terminate. It will only terminate if one of these inequalities is saturated. The operators which annihilate these states are then given in parenthesis behind the inequalities. Whether the inequalities are satisfied or not depends crucially on the range in which $\sss$ lies. Let us distinguish the following cases:
\begin{itemize}
\item[-] for $\sss\ge1$, the spectrum will be continuous,
\item[-] for $0<\sss<1$, there will be no states at all,
\item[-] and for $\sss\le0$, there will be a discrete series of eigenstates.
\end{itemize}
Let us start with the case $\sss\ge1$, and introduce a new quantum number $\scon\in\Rset_+$ such that $\sss=\scon^2+1$. The left hand sides of \eref{norm-cond} are now always positive, and we do not get any algebraic restrictions on the quantum numbers. But can we realize these states as square integrable functions on $\grpSL(2)$?  If we write the eigenstate as $\ket{\ttt,\mmm,\sss} = \ket{\ttt,\mmm;\psi}$, with a wave function $\psi(\euch)$, then we can study the behaviour of $\psi$ for $\euch\to\infty$. For large $\euch$, the eigenvalue equation of the Laplace Beltrami operator \eref{laplace} becomes 
\beq
  \expo{-2\euch} \deldel\euch \expo{2\euch} \deldel\euch \psi(\euch) 
                       \aln\sim  (\scon^2+1) \, \psi(\euch) \nwl \follows
  \psi(\euch) \aln\sim A \, \expo{- (1-\i\scon) \euch} 
               + B \, \expo{- (1+\i\scon) \euch}.
\eeq
We get a superposition of an ingoing and outgoing wave with wavelength $1/\scon$. They fall off with $\expt{-\euch}$. This is exactly the right behaviour for $\psi$ to become normalizable under the scalar product \eref{t-m-prod}, like usual plane waves in $L_2(\Rset)$. As $\euch$ is a radial coordinate, there must be a balance between ingoing and outgoing waves, and instead of two we only get one solution for $\psi$ for any fixed set of quantum numbers $\ttt$, $\mmm$ and $\sss$. So we have the following continuous series of eigenstates,
\beq
  \opr\SSS \Ket{\ttt,\mmm,\scon} = \hbar^2(\scon^2+1) \Ket{\ttt,\mmm,\scon} , 
  \qquad  \ttt,\mmm\in\Zset, \quad \scon \in \Rset_+.
\eeq
Next we consider the case $0<\sss<1$. This time we repace $\sss$ by a real number $\sdis$, such that $\sss=\sdis(2-\sdis)$, and for the moment we take $1<\sdis<2$. We then get some algebraic restrictions on the quantum numbers from \eref{norm-cond}. They are violated if $\ttt=\pm1$ or $\ttt+2\mmm=\pm1$. So we cannot have any states with $\ttt=\pm1$. But the inequalities are never saturated, and so we can change $\ttt$ by any multiple of two, using the $\TTT_\pm$ operators. This excludes all odd values for $\ttt$. On the other hand, if there is an eigenstate $\ket{\ttt,\mmm,\sss}$ for some even $\ttt$, then we can use the raising and lowering operations to arrive at a state $\ket{0,0,\sss}=\ket{0,0;\psi}$, and the wave function $\psi$ has to obey
\beq
  \psi''(\euch) + 2 \, \coth(2\euch) \, \psi'(\euch) = 
      \sdis (2-\sdis) \, \psi(\euch). 
\eeq
It is then not difficult to show that a solution to this differential equation can never be square integrable and smooth at the origin $\euch=0$. So there are no states with eigenvalues $0<\sss<1$.

Finally, let us study the case $\sss\le0$. Here we also use the parameter $\sdis$ such that $\sss=\sdis(2-\sdis)$, but now we take $\sdis\ge2$. Then we find the following condition on the quantum numbers,
\beq
   \ttt (\ttt + 2) \ge \sdis(\sdis-2)  \txt{and}  \ttt (\ttt - 2) \ge 
\sdis(\sdis-2).
\eeq
As $\sdis\ge2$, this is equivalent to the somewhat simpler condition
\beq
  \ttt \ge \sdis \txt{or} \ttt \le - \sdis,
\eeq
Now assume that $\sdis$ is \emph{not} an integer. Then these inequalities will never be saturated, which in turn means that starting from any state we can build up an infinite tower of states in both $\ttt$-directions, by acting on it with the $\TTT_\pm$ operators. Thereby, we can increase or decrease $\ttt$ by any multiple of $2$. However, at least one of these values will lie in the region between $\sdis$ and $-\sdis$, which is forbidden. So $\sdis$ has to be an integer. But even then we can create towers of states with elements falling into the forbidden region, unless the series terminates.

The only way to build up such a terminating tower is to start on the boundary, with quantum number $\ttt=\sdis$ or $\ttt=-\sdis$, and then act on this state with $\TTT_+$ or $\TTT_-$, respectively. Note that it is $T_-$ that annihilates the state at $\ttt=\sdis$ and vice verse, so what we get are two infinite towers, one with $\ttt=\sdis,\sdis+2,\sdis+4,\dots$ and the other one with $\ttt=-\sdis,-\sdis-2,-\sdis-4,\dots$. For a fixed $\ttt$, we therefore get the restriction that
\beq
  2 \le \sdis \le |\ttt|  \txt{and}  \ttt \equiv \sdis \, (\mod 2).
\eeq
There are no states for $\ttt=0$ and $\ttt=\pm1$. With a similar argument, we can determine the range of $\mmm$ for fixed $\sdis$ and $\ttt$. The result is  
\beq
  \mmm \ge \case12 (\sdis-\ttt) \txt{for} \ttt \ge 2, \qquad
  \mmm \le - \case12 (\sdis-\ttt) \txt{for} \ttt\le -2.
\eeq
Hence, for each value of $\sdis=2,3,4,\dots$, we have two ``ground states'' $\ket{\sdis, 0,\sdis}$ and $\ket{{-\sdis}, 0,\sdis}$, and all other states can be generated by acting with $\TTT_+$ and $\MMM_+$ onto $\ket{\sdis,0,\sdis}$, and similarly with the $\TTT_-$ and $\MMM_-$ onto $\ket{{-\sdis},0,\sdis}$. Note that thereby also negative values for $\mmm$ occur in the positive series and vice verse, because $\TTT_+$ \emph{decreases} $\mmm$ and $\TTT_-$ \emph{increases} it. Studying the asymptotic behaviour and of these states, we find that
\beq
       \psi(\euch) \sim A \, \expo{- \sdis \euch} 
                    + B \, \expo{ \sdis \euch}. 
\eeq
For this to be square integrable, we must have $B=0$. Thus, there is again only one state for each set of quantum numbers, and the whole discrete series of eigenstates is given by
\beq  
  \opr\SSS \Ket{\ttt,\mmm,\sdis} = \hbar^2 \, \sdis(2-\sdis) 
       \Ket{\ttt,\mmm,\sdis},  \qquad 
    \sdis = 2,3,4,\dots
\eeq
which splits into a positive and a negative series with
\beq
  \ttt \aln= \sdis,\sdis+2,\sdis+4, \dots, \qquad 
    \mmm \ge \case12(\sdis-\ttt) , \nwl
  \ttt \aln= -\sdis,-\sdis-2,-\sdis-4, \dots, \qquad 
    \mmm \le -\case12(\sdis-\ttt).
\eeq

\end{document}